\documentclass[
 aps,
 pra,
 showpacs,
 superscriptaddress,
 amsmath,amssymb,nofootinbib,
reprint,
 twocolumn
]{revtex4-1}

\usepackage[english]{babel}
\usepackage[T1]{fontenc}
\usepackage{hyperref}
\usepackage{graphicx}
\usepackage{subfigure}
\usepackage{amsmath}
\usepackage{amssymb}
\usepackage{slashed}
\usepackage{dsfont}
\usepackage{lipsum}

\newcommand{\gMq}{g_{\rm \pi qq}} 
\newcommand{\gMqSq}{g_{\rm \pi qq}^2} 
\newcommand{\dint}{\text{d}}
\renewcommand{\Re}{{\rm Re}\,}
\renewcommand{\Im}{{\rm Im}\,}
\newcommand{\Nc}{N_{\rm c}}
\newcommand{\Nf}{N_{\rm f}}
\newcommand{\Unit}{{\mathds 1}}
\newcommand{\psib}{\bar{\psi}}
\newcommand{\dslash}{\slashed{\partial}}
\newcommand{\Tr}{{\rm Tr}}
\renewcommand{\i}{{\rm i}}
\newcommand{\e}{{\rm e}}
\newcommand{\eps}{\varepsilon}
\renewcommand{\vec}[1]{\mbox{\boldmath $#1$}}
\newcommand{\wtEminus}{\widetilde{E}_-} 
\newcommand{\wtEplus}{\widetilde{E}_+} 

\newcommand{\ThermalInt}[2]{T\sum_{#1\in\mathds{Z}}\int\frac{\dint^3 #2}{(2\pi)^3}}

\begin{document}
\title{Shear Viscosities from Kubo Formalism in a large-{\boldmath $\Nc$} Nambu--Jona-Lasinio Model}

\author{Robert Lang}
  \email{robert.lang@ph.tum.de}
  \affiliation{Physik Department, Technische Universit\"{a}t M\"{u}nchen, D-85747 Garching, Germany}
\author{Norbert Kaiser}
  \email{nkaiser@ph.tum.de}
  \affiliation{Physik Department, Technische Universit\"{a}t M\"{u}nchen, D-85747 Garching, Germany}
\author{Wolfram Weise}
  \email{weise@tum.de}
  \affiliation{ECT*, Villa Tambosi, I-38123 Villazzano (TN), Italy}
  \affiliation{Physik Department, Technische Universit\"{a}t M\"{u}nchen, D-85747 Garching, Germany}

\date{October 19, 2015}

\begin{abstract}
In this work the shear viscosity of strongly interacting matter is calculated within a two-flavor Nambu--Jona-Lasinio model as a function of temperature and chemical potential. The general Kubo formula is applied, incorporating the full Dirac structure of the thermal quark spectral function and avoiding commonly used on-shell approximations. Mesonic fluctuations contributing via Fock diagrams provide the dominant dissipative processes. The resulting ratio $\eta/s$ (shear viscosity over entropy density) decreases with temperature and chemical potential. Interpolating between our NJL results at low temperatures and hard-thermal-loop results at high temperatures a minimum slightly above the AdS/CFT benchmark $\eta/s=1/4\pi$ is obtained.
\end{abstract}



\maketitle

\section{Introduction}
The quark-gluon plasma produced in heavy-ion collisions at RHIC and LHC is a hot and dense state of strongly correlated matter. It behaves like an almost-perfect fluid featuring a small ratio $\eta/s$ of shear viscosity to entropy density \cite{DuslingTeaney2008,LuzumGombeaudOllitrault2010,Song2013}. In this work we calculate viscous effects of interacting quarks within a two-flavor Nambu--Jona-Lasinio (NJL) model \cite{Nambu1961a,Nambu1961b,VoglWeise1991,Klevansky1992,HK1994,KLVW1990a,KLVW1990b,BuballaHabil2005}. A large-$\Nc$ scaling of the four-fermion vertex as inferred from QCD introduces a bookkeeping in which mesonic fluctuations (meson clouds around quarks) provide the dominant dissipative processes. The shear viscosity $\eta$ is calculated using the Kubo formalism \cite{Kubo57} similar as in Refs.~\cite{Alberico2008,HidakaKunihiro2011,HidakaKunihiro2011NJL,NamKao2013,GhoshLandau2013,Ghosh2014,Ghosh2014b,Christiansen2014}.
The new element of the present work is that the full Dirac structure of the thermal quark self-energy is included when evaluating the Kubo formula, thus avoiding commonly used approximations \cite{GhoshLandau2013,Fukutome2006,Fukutome2008Nucl,Fukutome2008Prog}. At the same time we extend our previous studies in Ref.~\cite{LangWeise:2014}.

The present paper is organized as follows: in Section \ref{StdNJL} we discuss the NJL model from the perspective of large-$\Nc$ scaling together with the gap equation and the Bethe-Salpeter equation. In addition we introduce the approximation scheme for the quark-meson coupling used in this work. In Section \ref{GeneralKubo} we develop the Kubo formula incorporating the full Dirac structure of the thermal quark self-energy. In Section \ref{CalcSelfEnergy} results for the quark self-energy $\Sigma_\beta$ generated by mesonic fluctuations are presented, assuming first an on-shell approximation. In this case the coupling between quarks and mesons is dissipative only at sufficiently high temperatures where pions can decay on-shell into quark-antiquark pairs and thus the quark self-energy receives an imaginary part.
In the next step constituent quarks off their mass shell are considered allowing for additional kinematic possibilities. Details of the calculation are displayed in the Appendices \ref{AppOnShell} and \ref{AppOffShell}. The results for the shear viscosity $\eta$ and the ratio $\eta/s$ are presented and discussed in Section \ref{Results}. In the high-temperature region gluonic degrees of freedom become dominant and results from hard-thermal loop calculations extend our NJL-model results \cite{Christiansen2014}. Finally, Section~\ref{Summary} gives a summary of our most important findings. 

\section{NJL model and mesonic clouds} \label{StdNJL}
In this work the simplest two-flavor NJL model is used, including scalar and pseudoscalar interactions only:
\begin{equation}
  \label{NJL2}
  \mathcal{L}=\psib \left(\i\dslash-\hat{m}_0\right)\psi+\frac G2\left[(\psib\psi)^2+(\psib \i\gamma_5\vec{\tau}\psi)^2\right],
\end{equation}
where $\psi = (u,d)^{\rm T}$ is the isospin-doublet quark field, $\hat{m}_0 = {\rm diag}(m_{\rm u},m_{\rm d})$ is the current-quark mass matrix (we work in the isospin limit, $m_{\rm u}= m_{\rm d} \equiv m_0$), and $\vec{\tau}$ denotes the vector of three isospin Pauli matrices. The effective four-fermion coupling $G$ is supposed to include non-perturbative (gluonic) dynamics. The large-$\Nc$ scaling of QCD implies $G\sim 1/\Nc$ in the NJL model. One should note that in QCD the color gauge symmetry is local whereas in the NJL model it  is reduced to a global symmetry.

\begin{figure}[t!]
\begin{center}
  \includegraphics[width=0.49\textwidth]{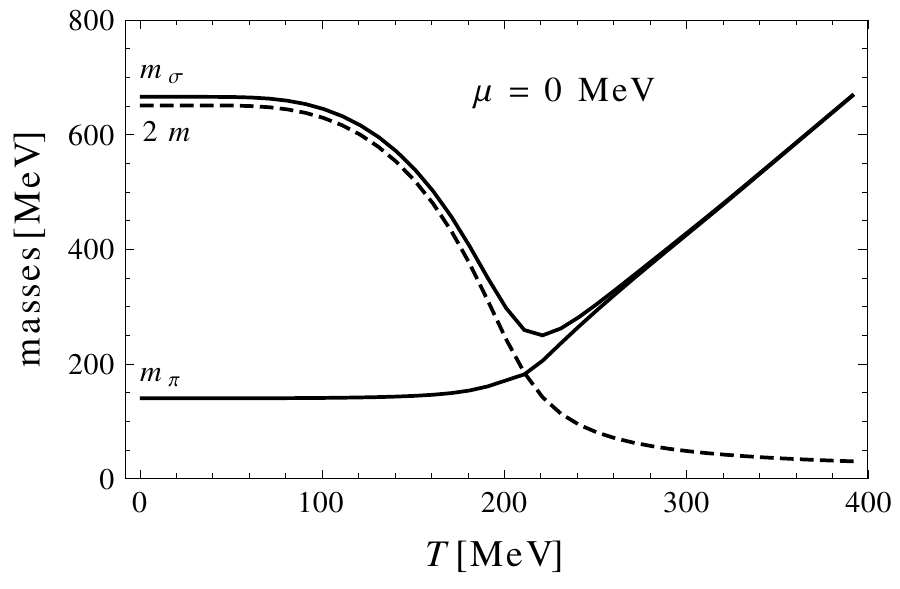}
\caption{Thermal masses of quarks (dashed curve) and mesons (solid curves) in the two-flavor NJL model with the parameter set given in Table~\ref{TableNJLquarkResults}.}
\label{Fig:ThermalMassesStandard}
\end{center}
\end{figure}

In this large-$\Nc$ counting a hierarchy of Dyson-Schwinger equations can be introduced \cite{QuackKlevansky1994,Buballa:2010}, where the leading order $\mathcal{O}(\Nc^0)$ is just the NJL gap equation in Hartree approximation:
\begin{equation}
  \label{GapHartree}
\begin{minipage}{0.5\textwidth}
\begin{minipage}{0.05\textwidth} ~ \end{minipage}
\begin{minipage}{0.20\textwidth}
\includegraphics[width=\textwidth]{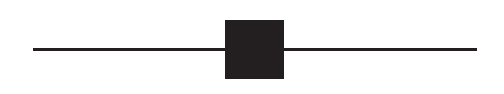}
\end{minipage}
\begin{minipage}{0.05\textwidth} \text{$\;=\;$} \end{minipage}
\begin{minipage}{0.26\textwidth}
\includegraphics[width=\textwidth]{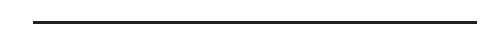}
\end{minipage}
\begin{minipage}{0.05\textwidth} \text{$+$} \end{minipage}
\begin{minipage}{0.20\textwidth}
\vspace{-0.5cm} \includegraphics[width=\textwidth]{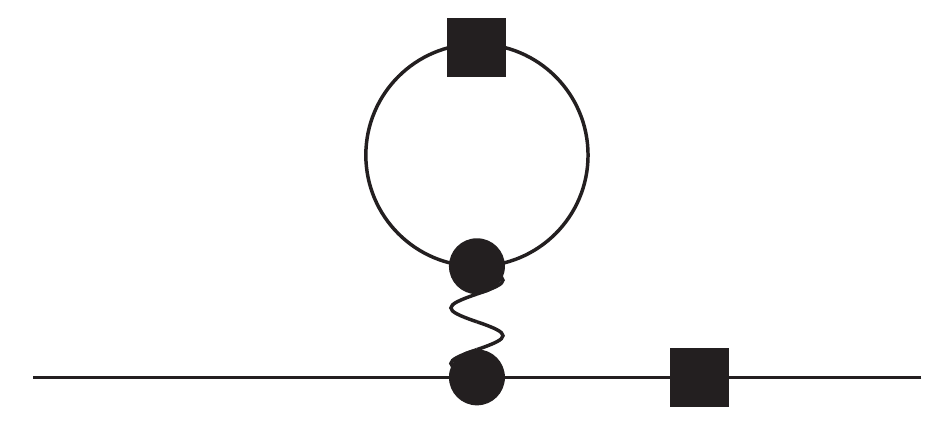}
\end{minipage}
\end{minipage}
\end{equation}
Lines with a black square denote full quark propagators whereas lines without denote bare quark propagators including the current-quark mass. In order to distinguish Hartree and Fock contributions generated by the four-quark vertex proportional to $G$ we have introduced the wavy line in the last diagram. It indicates that a color trace is involved in the quark loop. Hence, the diagrammatic gap equation \eqref{GapHartree} reads
\begin{equation}
\label{GapEquationHartree}
  m=m_0-G\langle\psib\psi\rangle\;.
\end{equation}
It includes contributions from explicit chiral symmetry breaking, $m_0$, and from the quark condensate:
\begin{equation}
\label{DefChiralCond}
  \langle\psib\psi\rangle=-\frac{2\Nc}{\pi^2}\int_0^\Lambda\dint p\;\frac{p^2\, m}{E_p} \left[1-n^+_{\rm F}(E_p)-n^-_{\rm F}(E_p)\right],
\end{equation}
where $n_{\rm F}^\pm(E_p)=\left[1+\exp\left(\beta(E_p\mp\mu)\right)\right]^{-1}$ denotes the Fermi distribution functions with $E_p=\sqrt{p^2+m^2}$, and $\beta=1/T$ is the inverse temperature. In the commonly used mean-field approximation, the (thermal) constituent-quark mass is determined by solving just this Hartree part of the gap equation. The resulting quark mass within this approximation is shown in Fig.~\ref{Fig:ThermalMassesStandard}.

At next-to-leading order the mesonic modes are obtained from the well-known Bethe-Salpeter equation (BSE) in random-phase approximation:
\begin{equation}
  \label{BSEderivedLargeNc}
\begin{minipage}{0.49\textwidth}
\begin{minipage}{0.27\textwidth}
\includegraphics[width=\textwidth]{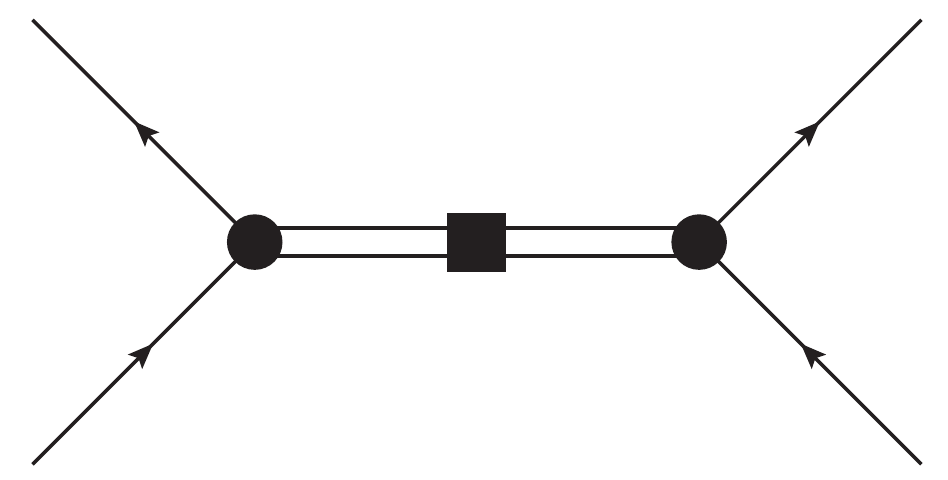}
\end{minipage}
\begin{minipage}{0.05\textwidth} \text{$\;=\;$} \end{minipage}
\begin{minipage}{0.17\textwidth}
\includegraphics[width=\textwidth]{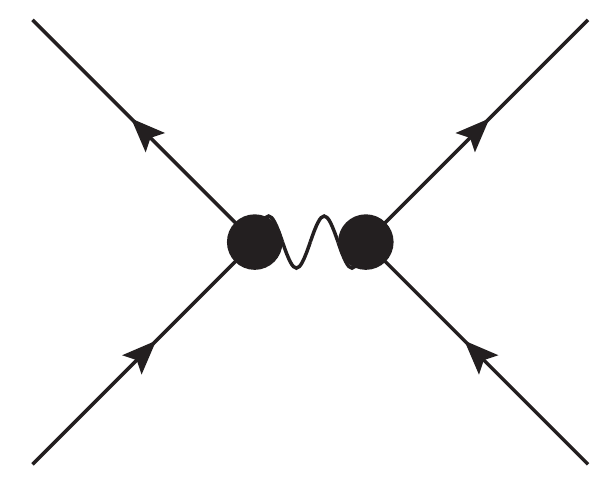}
\end{minipage}
\begin{minipage}{0.05\textwidth} \text{$+$} \end{minipage}
\begin{minipage}{0.4\textwidth}
\includegraphics[width=\textwidth]{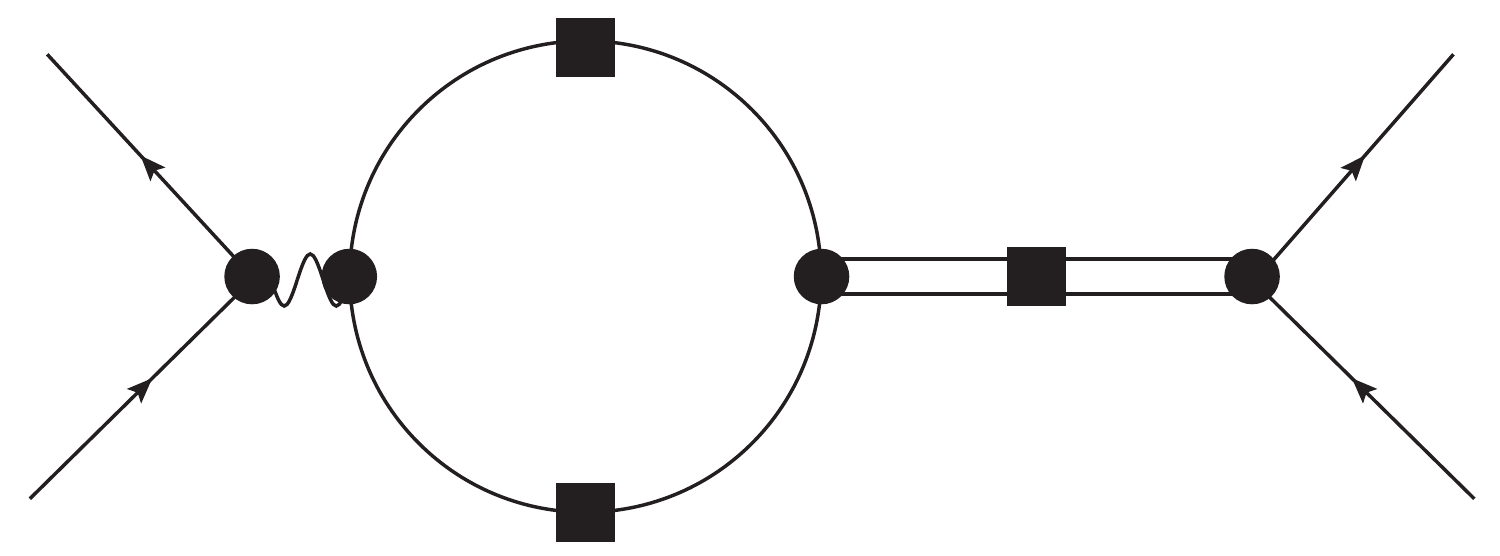}
\end{minipage}
\end{minipage}
\end{equation}

When going beyond the mean-field approximation, the NJL gap equation includes Fock terms that are suppressed by $1/\Nc$:
\begin{equation}
\begin{aligned}
  \label{GapDerivedLargeNc} ~ \\
\begin{minipage}{0.49\textwidth}
\begin{minipage}{0.15\textwidth} \includegraphics[width=\textwidth]{DSEFermion.pdf}
\end{minipage}
\begin{minipage}{0.04\textwidth} \text{$=$} \end{minipage}
\begin{minipage}{0.15\textwidth} \includegraphics[width=\textwidth]{DSEFermionBare.pdf}
\end{minipage}
\begin{minipage}{0.04\textwidth} \text{$+$} \end{minipage}
\begin{minipage}{0.20\textwidth}
\vspace{-0.5cm} \includegraphics[width=\textwidth]{DSEFermionHartree.pdf}
\end{minipage}
\begin{minipage}{0.04\textwidth} \text{$+$} \end{minipage}
\begin{minipage}{0.25\textwidth}
\vspace{-0.3cm} \includegraphics[width=\textwidth]{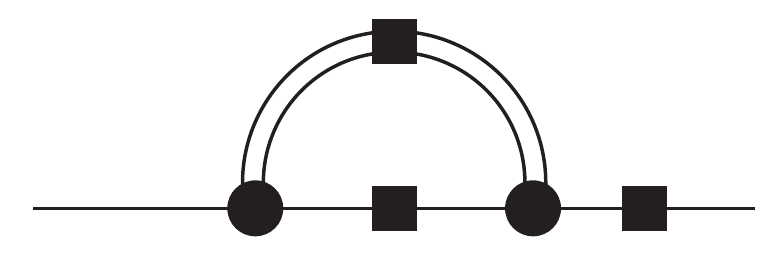}
\end{minipage}
\end{minipage}
\end{aligned}
\end{equation}
A self-consistent treatment of the gap equation including Fock terms with mesonic modes and the Bethe-Salpeter equation describing these mesonic modes is approximated by a common procedure \cite{QuackKlevansky1994,HeckmannBuballaWambach2012}: the mesonic fluctuations are evaluated using the Hartree solution of the gap equation only which ensures a consistent large-$\Nc$ counting of the NJL model up to next-to-leading order $1/\Nc$.

The last (Fock) diagram in the extended gap equation \eqref{GapDerivedLargeNc} represents the mesonic clouds which couple mesonic fluctuations to constituent quarks. This coupling is described by Yukawa interactions with a single quark-meson coupling, $g_{\rm\pi qq}$, implied by chiral symmetry:
\begin{equation}
  \Delta\mathcal{L}=-g_{\rm \pi qq}\left(\psib\i\gamma_5\vec{\tau}\psi\cdot\vec{\pi} + \psib\sigma\psi\right).
\end{equation}
Solving the Bethe-Salpeter equation \eqref{BSEderivedLargeNc} gives (renormalized) meson propagators from which meson masses can be extracted. The resummation of quark-antiquark scattering modes leads to meson propagators for any of the pions (${\rm P}$) or the sigma boson (${\rm S}$):
\begin{equation}
\label{RPAresummed}
  D_{\rm M}=G+G\Pi^{\rm S/P} D_{\rm M}=\frac{G}{1-\Pi^{\rm S/P} G}\;.
\end{equation}
Here we have introduced the polarization tensors $\Pi^{\rm S/P}$ corresponding to the scalar or pseudoscalar quark-antiquark loops. They have the form
\begin{equation}
\label{QuarkAntiquarlLoopResult}
  \Pi^{\rm S/P}(\vec{p},\omega_n)=8\Nc I_1+4\Nc N^{\rm S/P} I_2(\vec{p},\omega_n)\;,
\end{equation}
where $N^{\rm P}=-\left(\omega_n^2+\vec{p}^2\right)$ and $N^{\rm S}=N^{\rm P}-4m^2$ refer to the pion and sigma modes, respectively. Explicit expressions for $I_1$ and $I_2(\vec{p},\omega_n)$ can be found in Appendix~\ref{AppBSE}. Note that the gap equation in Hartree approximation \eqref{GapEquationHartree} involves also the loop integral $I_1$:
\begin{equation}
\label{ThermalGapEquation}
  m=m_0+8G\Nc mI_1\;.
\end{equation}
\begin{table}[t!]
\begin{center}
   \resizebox{0.5\textwidth}{!}{\begin{tabular}{|c|c|c||c|c|c|c|} \hline
\multicolumn{3}{|c||}{Input} & \multicolumn{4}{c|}{Output $[{\rm MeV}]$} \\ \hline
\hline
\parbox[0pt][0.85cm][c]{0cm}{} $m_0$ & $G$ & $\Lambda$ &
$\hspace{0.2cm} m \hspace{0.2cm}$ & 
$\hspace{0.2cm} m_\pi \hspace{0.2cm}$ &
$\hspace{0.2cm} f_\pi \hspace{0.2cm}$ & 
$\hspace{0.1cm} \langle\psib\psi\rangle^{1/3} \hspace{0.1cm}$  \\ \hline 
\parbox[0pt][0.85cm][c]{0cm}{} $5.50\;{\rm MeV}$ & $10.1\;{\rm GeV}^{-2}$ & $651\;{\rm MeV}$ & $325$ & $140$ & $92.4$ & $-316$ \\ \hline
  \end{tabular}}
\end{center}
\caption{NJL parameter set and resulting physical quantities}
\label{TableNJLquarkResults}
\end{table} 
Poles of the meson propagators $D_{\rm M}$ can appear only in Minkowski space, therefore one performs the analytical continuation $\i\omega_n\mapsto \omega+\i\eps$. In general the polarization tensor is complex which is evident at high temperatures where a mesonic resonance instead of a bound state is realized. Therefore, we define the meson mass $m_{\rm M}$ at $\vec{p}=\vec{0}$ as a solution of\footnote{In principle, there is a momentum dependence of the pion mass. We have chosen to define masses always in the reference frame of the heat bath, so $\vec{p}=\vec{0}$ and $\omega_n^2=-m_\pi^2$.}:
\begin{equation}
\label{MesonPoleMassDefinition}
  \Re D_{\rm M}^{-1}(\vec{0},-\i m_{\rm M})=0\;.
\end{equation}
In the pseudoscalar channel the (renormalized) pion propagator reads
\begin{equation}
\label{RescaledPionPropDPi}
  D_\pi(\vec{p},\omega_n)=\frac{G}{\frac{m_0}{m}+4G\Nc (\omega_n^2+\vec{p}^2)I_2(\vec{p},\omega_n)}\;,
\end{equation}
from which the pion mass can be calculated by solving
\begin{equation}
\label{MassPion}
  m_\pi^2=\frac{m_0}{m}\frac{1}{4G\Nc\;\Re I_2(\vec{0},-\i m_\pi)}\;.
\end{equation}
If one considers the scalar channel instead, the mass of the sigma boson can be extracted\footnote{Note that this relation between $m_\sigma$ and $m_\pi$ is valid only if the energy dependence of $I_2$ is negligible. For high $T$ pions and the sigma boson are degenerate and one has $I_2(-\i m_\pi)\approx I_2(-\i m_\sigma)$.}:
\begin{equation}
\label{MassSigma}
  m_\sigma^2=m_\pi^2+4m^2\;.
\end{equation}
At high temperatures $T$, far above the chiral transition temperature, the scalar and pseudoscalar modes tend to degenerate: $m_\sigma^2\to m_\pi^2$. This goes along with the restoration of chiral symmetry. The results for the thermal meson masses are also shown in Fig.~\ref{Fig:ThermalMassesStandard}. The parameter set which has been used to calculate $m(T)$, $m_\pi(T)$ and $m_\sigma(T)$ is given in Table~\ref{TableNJLquarkResults}. It reproduces physical values for $m_\pi=140\;{\rm MeV}$, for the pion decay constant $f_\pi=92.4\;{\rm MeV}$, and realistic values for the constituent-quark mass $m=325\;{\rm MeV}$ and the chiral condensate $\langle\psib\psi\rangle=-(316\;{\rm MeV})^3$. Due to the absence of confinement in the NJL model, the thermal pion can decay on-shell into two thermal constituent quarks. The critical temperature for this is called \textit{Mott temperature} $T_{\rm M}$ and determined by $m_\pi(T_{\rm M})=2m(T_{\rm M})$. At vanishing chemical potential it has the value $T_{\rm M}\approx 212\;
{\rm MeV}$.
Therefore, for the shear viscosity arising from mesonic fluctuations, on-shell dissipative effects are possible only for $T>T_{
\rm M}$.

\begin{figure}[t!]
\begin{center}
  \includegraphics[width=0.49\textwidth]{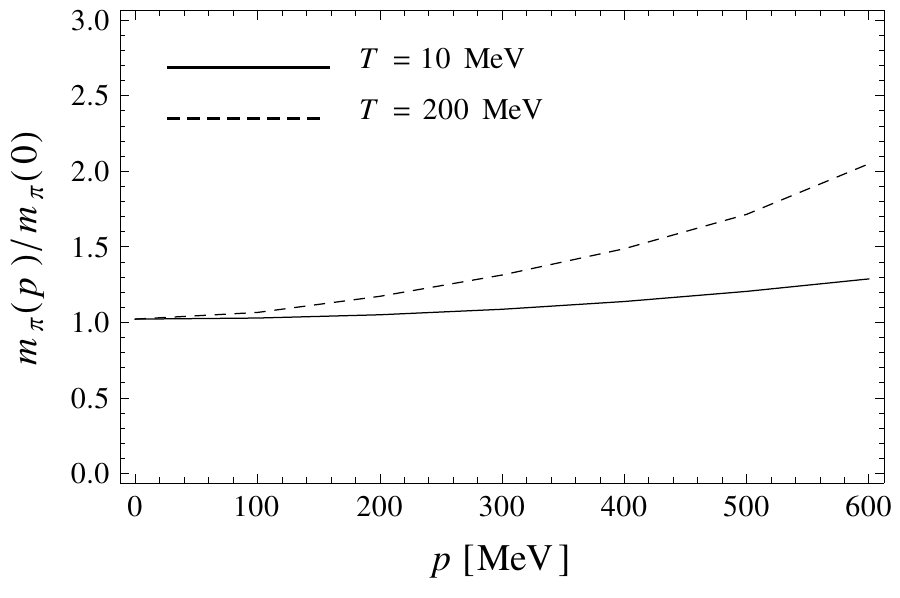}
\caption{Momentum dependence of the pion mass at two different temperatures}
\label{Fig:PionMassPdep}
\end{center}
\end{figure}
We continue with examining the validity of the standard pole approximation for determining meson masses. In Minkowski space, the solution of the BSE (at finite $T$ and $\mu$) reads for the pionic mode:
\begin{equation}
\label{AppDPiProp}
  D_\pi(\vec{p},-\i\omega)=\frac{G}{\frac{m_0}{m}-4G\Nc\left(\omega^2-\vec{p}^2\right)I_2(\vec{p},-\i\omega)}\;.  
\end{equation}
The fact that $I_2(\vec{p},-\i\omega)$ is energy and momentum dependent implies that the standard pole-mass approximation,
\begin{equation}
\label{PionPropPoleCoupling}
  \frac{-g_{\rm \pi qq,static}^2}{\omega^2-\vec{p^2}-m_\pi^2+\i\eps}\;,
\end{equation}
does not reproduce the full $(\vec{p},\omega)$ dependence of the propagator. A general quark-pion coupling can be introduced by
\begin{equation}
  g^2_{\rm\pi qq}(\omega,\vec{p})=-\left(\omega^2-\vec{p}^2-m^2_\pi(\vec{p})\right)D_\pi(\vec{p},-\i\omega)\;,
\end{equation}
with the momentum dependent pion mass $m_\pi(\vec{p})$ defined as a solution of
\begin{equation}
  \Re D_\pi^{-1}(\vec{p},-\i m_\pi(\vec{p}))=0\;.
\end{equation}
The results for the momentum-dependent pion mass $m_\pi(\vec{p})$ are shown in Fig.~\ref{Fig:PionMassPdep}. The pion becomes more massive when it carries additional momentum. This qualitative behavior of $m_\pi(\vec{p})$ is consistent with the fact that the constituent-quark mass $m(p)$ decreases as function of momentum. The associated tendency towards chiral symmetry restoration weakens the Goldstone boson character of the pion.

\begin{figure}[t!]
\begin{center}
  \subfigure{\includegraphics[width=0.49\textwidth]{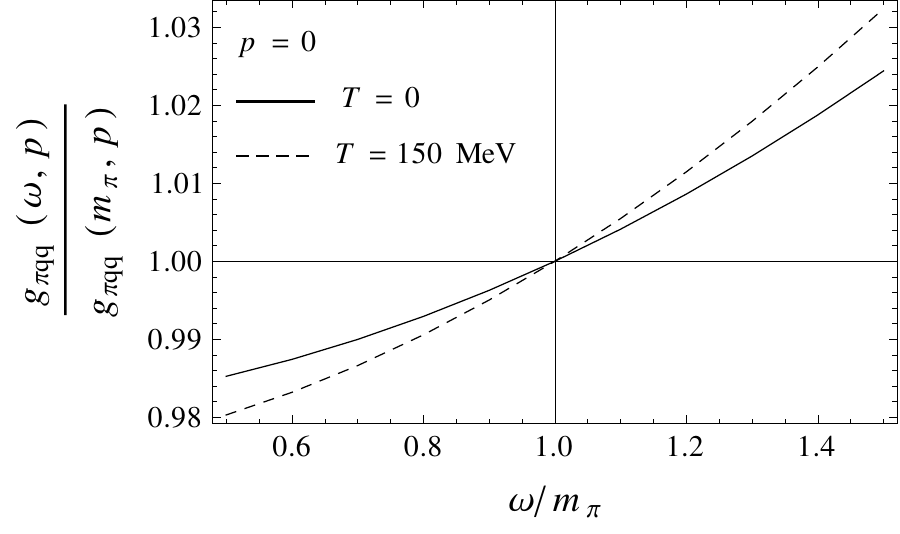}}
  \subfigure{\includegraphics[width=0.49\textwidth]{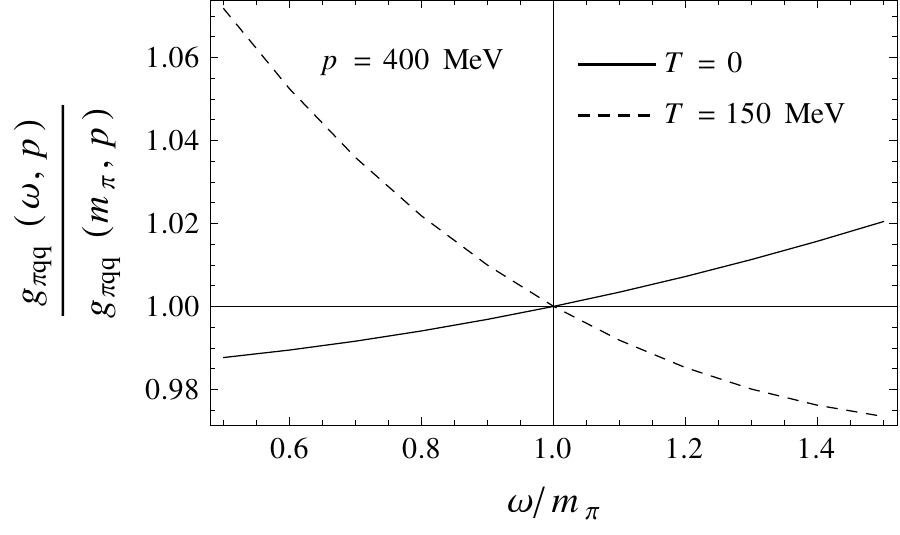}}
\caption{Corrections to the quark-pion coupling in a $50\%$ range around the pole position. (a) Vanishing pion momentum; (b) pion momentum $p=p=400\;{\rm MeV}$}
\label{Fig:CheckPoleMass}
\end{center}
\end{figure}

As usual we define the quark-meson coupling as the residue of the full meson propagator at vanishing momentum \cite{HatsudaKunihiro:1987}:
\begin{equation}
   g_{\rm\pi qq}^{-2}=-\left.\frac{\dint}{\dint\omega^2}D^{-1}_{\rm \pi}(\vec{0},-\i\omega)\right|_{\omega^2=m_{\rm \pi}^2}\;.
\end{equation}
From the pion propagator $D_\pi$ in Eq.~\eqref{AppDPiProp} we get immediately
\begin{equation}
\begin{aligned}
\label{QuarkMesonCouplingDerivateResult}
  g_{\rm\pi qq}^{-2}&=4\Nc\left.\left(I_2(-\i\omega)+\omega^2\frac{\dint I_2}{\dint\omega^2}\right)\right|_{\omega^2=m_{\rm \pi}^2}=\\
  &=g_{\rm\pi qq,static}^{-2}\left(1+\frac{\omega^2}{I_2(-\i\omega)}\left.\frac{\dint I_2}{\dint\omega^2}\right|_{\omega^2=m_{\rm \pi}^2}\right).
\end{aligned}
\end{equation}
We have identified the static quark-meson coupling, $g_{\rm\pi qq,static}$, where the energy dependence of $I_2(\vec{0},-\i\omega)$ is neglected and find $g_{\rm\pi qq,static}^{-2}=4\Nc I_2(\vec{0},-\i m_\pi)$, cf. Eq.~\eqref{PionPropPoleCoupling}. In Fig.~\ref{Fig:CheckPoleMass} we compare the two approaches for calculating the quark-pion coupling. When staying in a $50\%$ interval around the pion pole, $0.5m_\pi<\omega<1.5m_\pi$, we find indeed that the usual treatment, fixing $\gMq$ at the pion pole, is a rather good approximation. It is interesting to note that for vanishing three-momentum at $T=150\;{\rm MeV}$ the approximated quark-pion coupling, $g_{\rm\pi qq,static}$ overestimates the actual coupling when evaluating at energies smaller than the pion mass. In contrast, for high momenta but the same temperature, the coupling is underestimated by up to $6\%$ which is still a good approximation. Only at very high temperatures and far away from the actual pole mass sizable deviations occur. 
However, apart from such extreme values, one can conclude that the approximated quark-pion coupling is acceptable and corrections beyond Eq.~\eqref{QuarkMesonCouplingDerivateResult} contribute at the order of only a few percent. Therefore, we do not take any momentum dependence into account apart from the derivative correction in Eq.~\eqref{QuarkMesonCouplingDerivateResult}. This leads to a simplified treatment of the mesonic fluctuations since the quark-meson coupling is just a constant and therefore does not affect the momentum integration.

\section{Kubo formalism and quark self-energy} \label{GeneralKubo}
In the Kubo formalism the shear viscosity is related to a correlator of the energy-momentum tensor. Assuming an infinite homogeneous medium close to thermal equilibrium, the (frequency dependent) shear viscosity is given by \cite{LangWeise:2014}:
\begin{equation}
  \label{KuboEta}
\eta(\omega)=\frac{\beta}{15}\int_0^\infty\dint t\;\e^{\i\omega t}\int\dint^3x\; \left(T_{\mu\nu}(t,\vec{x}),T^{\mu\nu}(0,\vec{0})\right),
\end{equation}
with the energy-momentum tensor of quarks:
\begin{equation}
  \label{TmunuNJLmodel}
T_{\mu\nu} =\frac{\partial\mathcal{L}}{\partial(\partial^\mu\psi)}\,\partial_\nu\psi -g_{\mu\nu}\mathcal{L} =\i\psib\gamma_\mu\partial_\nu\psi -g_{\mu\nu}\mathcal{L}\;.
\end{equation}
The correlator in the integrand of the Kubo formula \eqref{KuboEta} is defined through the thermal expectation values with
\begin{equation}
\label{DefCorrelatorXY}
  (X,Y)=\frac 1\beta\int_0^\beta\dint\xi\;\langle X\e^{-\xi H}Y\e^{\xi H}\rangle_0\;,
\end{equation}
with $H$ the Hamiltonian. As discussed in our previous work \cite{LangWeise:2014}, the shear viscosity can be written in terms of the quark spectral function \text{$\rho=-\frac{1}{\pi}\,\Im G_{\rm R}$} (with $G_{\rm R}$ the retarded quark propagator, see also Refs.~\cite{Fukutome2006,Fukutome2008Nucl,Fukutome2008Prog}):
\begin{equation}
\begin{aligned}
\label{EtaLeadingNcAsTwoSpectralDensities}
  \left.\eta\right|_{\omega=0} =\;&\frac{\pi}{T} \int_{-\infty}^\infty \dint\eps\int\frac{\dint^3 p}{(2\pi)^3}\;p_x^2\, n^+_{\rm F}(\eps)\big(1-n^+_{\rm F}(\eps)\big)\,\\
  & \times \Tr\left[\gamma_2\,\rho(\eps,\vec{p})\,\gamma_2\,\rho(\eps,\vec{p})\right].
\end{aligned}
\end{equation}
In Ref.~\cite{LangWeise:2014} the shear viscosity $\eta[\Gamma(p)]$ has been explored assuming a simple parameterization of the thermal quark propagator with a schematic (momentum dependent) spectral width, $\Gamma(p;T,\mu)$:
\begin{equation}
\label{KuboQuasiPartApprQuarkProp}
  G_{\rm R}(p_0,\vec{p})=\frac{1}{\slashed{p}-m+\i\,{\rm sgn}(p_0)\Gamma(p)}\;,
\end{equation}
from which the quark spectral function can be derived. The general Dirac structure is, however, richer than the parameterization in Eq.~\eqref{KuboQuasiPartApprQuarkProp}. Due to the breaking of Lorentz invariance in the thermal medium, three functions $A,B,C$ are necessary to specify the quark spectral function:
\begin{equation}
\label{DefABCD}
  \rho(p_0,\vec{p})=-\frac{1}{\pi D}\left[mA+p_0\gamma_0B-\vec{p}\cdot\vec{\gamma}\,C\right],
\end{equation}
with a denominator $D$. These four functions depend on the (off-shell) energy $p_0$, the three-momentum $\vec{p}$, and the thermal parameters $T$ and $\mu$. They can be determined from the thermal quark propagator in Minkowski space
\begin{equation}
\begin{aligned}
  G_{\rm R}(p_0,\vec{p}) &=\frac{1}{\slashed{p}-m-\hat{\Sigma}}\\
  &\hspace{-1.5cm}=\frac{m(1+\hat{\Sigma}_0)+p_0\gamma_0(1+\hat{\Sigma}_4)-\vec{p}\cdot\vec{\gamma}(1+\hat{\Sigma}_3)}{p_0^2(1+\hat{\Sigma}_4)^2-\vec{p}^2(1+\hat{\Sigma}_3)^2-m^2(1+\hat{\Sigma}_0)^2}\;.
\end{aligned}
\end{equation}
The general Dirac structure of the thermal quark self-energy is
\begin{equation}
  \hat{\Sigma}=m\hat{\Sigma}_0+\vec{p}\cdot\vec{\gamma}\hat{\Sigma}_3-p_0\gamma_0\hat{\Sigma}_4\;,
\end{equation}
with three dimensionless functions $\hat{\Sigma}_j(p_0,\vec{p})$. Incorporating mesonic fluctuations within the NJL model, the self-energy $\hat{\Sigma}$ receives contributions from three pions and one sigma boson:
\begin{equation}
\label{DefSigmanTot}
  \hat{\Sigma}_j=3\Sigma_j^{\rm P}+\Sigma_j^{\rm S}\;, \;\;\; \mbox{for}\;\;\; j=0,3,4\;,
\end{equation}
with $\Sigma_j^{\rm S/P}$ further specified after Eq.~\eqref{ThermalParameterizationSelfEnergy}.

For the present calculation we take into account only the relevant imaginary parts,
\begin{equation}
  \Im \hat{\Sigma}_j=\rho_j\;.
\end{equation}
In doing so, we ignore the momentum dependence of the constituent-quark mass as it arises from mesonic Fock contributions. Formally, this approximation is equivalent to readjusting the NJL parameters and introducing a new set $(m_0,G,\Lambda)^{\rm new}$ that will depend on the thermal variables $T$ and $\mu$, and on energy and momentum. The resulting thermal quark propagator is:
\begin{equation}
  G_{\rm R}(p_0,\vec{p}) = \frac{m(1+\i\rho_0)+p_0\gamma_0(1+\i\rho_4)-\vec{p}\cdot\vec{\gamma}(1+\i\rho_3)}{N_1+2\i\,N_2}\;,
\end{equation}
with the two auxiliary functions $N_{1,2}(p_0,p)$:
\begin{equation}
\begin{aligned}
\label{DefAuxiliaryFctN1N2}
  N_1 &= p_0^2(1-\rho_4^2)-p^2(1-\rho_3^2)-m^2(1-\rho_0^2)\;,\\
  N_2 &= p_0^2\rho_4-p^2\rho_3-m^2\rho_0\;.
\end{aligned}
\end{equation}
The four energy and momentum-dependent functions $A,B,C,D$ parameterizing the quark spectral function $\rho$ in Eq.~\eqref{DefABCD} are thus identified as:
\begin{equation}
\label{ResultsABCD}
\begin{aligned}
  A &=\rho_0 N_1-2N_2\;, \;\;\;\;\; B =\rho_4 N_1-2N_2\;, \\
  C &=\rho_3 N_1-2N_2\;, \;\;\;\;\; D =N_1^2+4N_2^2\;.
\end{aligned}
\end{equation}
The evaluation of the shear viscosity is now reduced to carrying out all the traces for the integrand of  Eq.~\eqref{EtaLeadingNcAsTwoSpectralDensities}:
\begin{equation}
  \Tr [\gamma_2\rho\gamma_2\rho] =\frac{4\Nc\Nf}{\pi^2D^2}\left[-m^2A^2+p_0^2B^2-p^2C^2+2p_y^2C^2\right].
\end{equation}
After angular integration the shear viscosity reads:
\begin{equation}
  \label{GeneralDiracEtaCombinedFULL}
\begin{aligned}
  \eta&=\frac{2\Nc\Nf}{3\pi^3T}\int_{-\infty}^\infty\dint\epsilon\int_0^\Lambda\dint p\; n_{\rm F}^+(\epsilon)(1-n_{\rm F}^+(\epsilon))\\
  &\times\frac{p^4}{D^2(\epsilon,p)}\left[-m^2A^2(\epsilon,p)-\frac{3}{5}p^2C^2(\epsilon,p)+\epsilon^2B^2(\epsilon,p)\right],
\end{aligned}
\end{equation}
with $\Lambda=651\;{\rm MeV}$ the NJL cutoff and $\Nc\Nf=6$. Quite remarkably, negative and positive contributions balance to an overall positive shear viscosity $\eta>0$. It is important to note that according to the representation in Eq.~\eqref{GeneralDiracEtaCombinedFULL}, $\eta$ is an even function of the chemical potential $\mu$. This is ensured by a separate integration over positive and negative energies and the property $I(-\epsilon,\mu)=I(\epsilon,-\mu)$ of the entire integrand.

\section{Quark self-energy from mesonic fluctuations} \label{CalcSelfEnergy}
\begin{figure}[t!]
\begin{center}
  \includegraphics[width=0.48\textwidth]{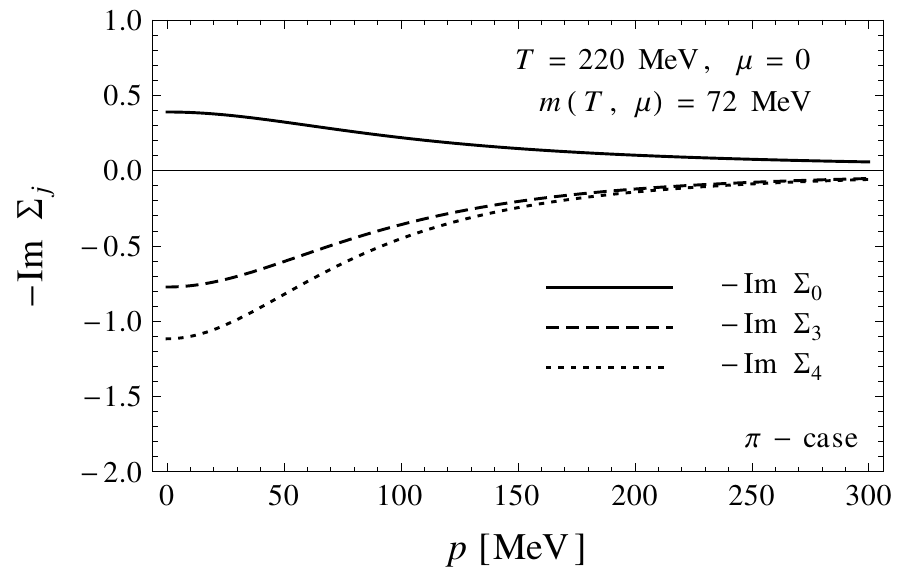}
  \includegraphics[width=0.48\textwidth]{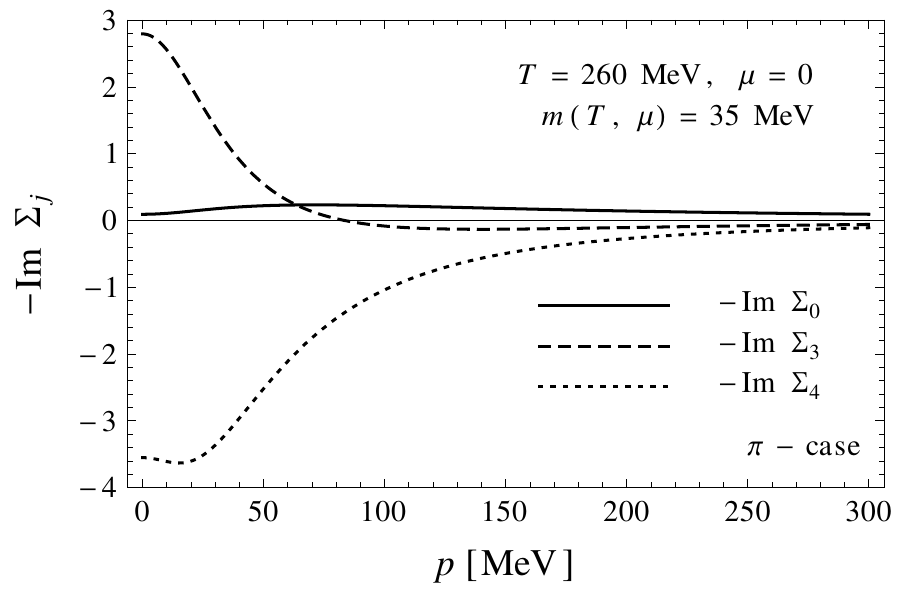}
  \caption{The (negative) imaginary parts of the quark self-energy contributions $\Sigma_j$, $j=0,3,4$, from mesonic fluctuations. They have been defined in Eq.~\eqref{ThermalParameterizationSelfEnergy}.}
  \label{Fig:ImaginaryPartsOfSigmas}
\end{center}
\end{figure}

In this section we evaluate the quark self-energy arising from the Fock diagram with mesonic fluctuations \eqref{GapDerivedLargeNc}. They introduce non-vanishing imaginary parts $\rho_j$ at next-to-leading order in $1/\Nc$.
\subsection{On-shell quarks}
Let us first consider quarks with on-shell kinematics $p_0^2=m^2+\vec{p}^2$. The Matsubara frequencies for in-medium quarks are $\nu_n=(2n+1)\pi T-\i\mu$. Note that the frequencies for antiquarks are $\nu_n^*=(2n+1)\pi T+\i\mu$. There are $N_{\rm f}^2-1=3$ equal contributions from the pseudoscalar channel (pions, $\Gamma^{\rm P}=\i\gamma_5$) and one contribution from the scalar channel (sigma boson, $\Gamma^{\rm S}=\Unit$). The corresponding self-energies are calculated as:
\begin{equation}
\begin{aligned}
\label{FockDiagramSigmaBetaSP}
  \Sigma^{\rm S/P}_\beta(\vec{p},\nu_n)\; &= \includegraphics[width=0.15\textwidth]{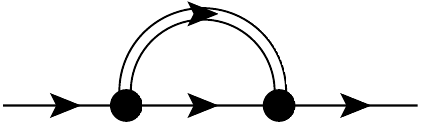} \\
  &\hspace{-2cm}=g_{\rm\pi qq}^2\ThermalInt{m}{q}\,\Gamma^{\rm S/P}G_\beta^{\rm F}(\vec{q},\nu_m) \\
  &\hspace{1.2cm}\times \Gamma^{\rm S/P}G_\beta^{\rm B}(\vec{p}-\vec{q},\nu_n-\nu_m)\;,
\end{aligned}
\end{equation}
with the thermal quark and meson propagators,
\begin{equation}
\begin{aligned}
  G_\beta^{\rm F}(\vec{p},\nu_n) &=\frac{\nu_n\gamma_4-\vec{p}\cdot\vec{\gamma}+m}{\nu_n^2+\vec{p}^2+m^2}\;,\\
  G_\beta^{\rm B}(\vec{p},\omega_n) &=\frac{1}{\omega_n^2+\vec{p^2}+m^2}\;,
\end{aligned}
\end{equation}
where $\omega_n=2n\pi T$ are bosonic Matsubara frequencies. Its Dirac structure has the following form:
\begin{equation}
  \label{ThermalParameterizationSelfEnergy}
\Sigma_{\beta}^{\rm S/P}(\vec{p},\nu_n) = \pm m\,\Sigma_0-\vec{p}\cdot\vec{\gamma}\,\Sigma_3+\nu_n\gamma_4\,\Sigma_4\;,
\end{equation}
with three dimensionless functions $\Sigma_j(p,\nu_n)$. The plus and minus sign in front of $\Sigma_0$ refers to the (scalar) sigma boson and the (pseudoscalar) pion, respectively: $\Sigma^{\rm S/P}_{3,4}=\Sigma_{3,4}$ but $\Sigma^{\rm S/P}_0=\pm\Sigma_0$. We note that in the single-width approximation in Eq.~\eqref{KuboQuasiPartApprQuarkProp} would give $\Im\Sigma_0=-\frac 1m \Gamma(p)$ and $\Sigma_{3,4}=0$.

We now analytically continue the quark self-energy to Minkowski space, $\nu_n\mapsto -\i p_0$, and extract the imaginary parts of $\Sigma_j$ relevant for calculating the shear viscosity. The detailed derivation can be found in Appendix \ref{AppOnShell}. Here we only state the results ($j=0,3,4$):
\begin{equation}
  \label{SummaryOnShellImaginaryParts}
\begin{aligned}
  &\Im\,\Sigma_j(p,-\i p_0)\\
  &=-\frac{\gMqSq}{16\pi p}\int_{E_{\rm min}}^{E_{\rm max}}\dint E_f\,\mathcal{F}_j\left[n_{\rm B}(E_b)+n_{\rm F}^-(E_f)\right],
\end{aligned}
\end{equation}
with $E_b=E_f+p_0$, $p_0=\sqrt{m^2+p^2}$, the weight factors
\begin{equation}
\begin{aligned}
\label{DefinitionF034}
  \mathcal{F}_0 &= 1\;,\\
  \mathcal{F}_3 &= \frac{m_{\rm M}^2-2m^2-2E_fp_0}{2p^2}\;,\\
  \mathcal{F}_4 &= -\frac{E_f}{p_0}\;,
\end{aligned}
\end{equation}
and the integration boundaries
\begin{equation}
\begin{aligned}
\label{EminmaxText}
  E_{{\rm max},{\rm min}} &=\frac{1}{2m^2}\left[\left(m_{\rm M}^2-2m^2\right)\sqrt{m^2+p^2}\right.\\
  &\hspace{1.5cm}\left.\pm p\,m_{\rm M}\sqrt{m_{\rm M}^2-4m^2}\right].
\end{aligned}
\end{equation}
The index ${\rm M}$ is either ${\rm S}$ (sigma boson) or ${\rm P}$ (pion). The remaining integral over $E_f$ in Eq.~\eqref{SummaryOnShellImaginaryParts} can be performed and one finds the following analytical expressions: 
\begin{equation}
  \label{FinalAnalytResultImSigma0}
\Im\,\Sigma_0(p,-\i p_0)=\frac{\gMqSq}{16\pi p}\,T\ln\frac{n_{\rm F}^-(E_{\rm min})\,n_{\rm B}(E_{\rm max}+p_0)}{n_{\rm F}^-(E_{\rm max})\,n_{\rm B}(E_{\rm min}+p_0)}\;,
\end{equation}
with $n_{\rm F}^-$ denoting the antiquark distribution function defined after Eq.~\eqref{DefChiralCond} and $n_{\rm B}(E)=\left[\exp(\beta E)-1\right]^{-1}$ the Bose distribution function.

The representations of $\Sigma_3$ and $\Sigma_4$ contain energy-dependent prefactors, $\mathcal{F}_3$ and $\mathcal{F}_4$, respectively, which lead to more complex results. After introducing the auxiliary function
\begin{equation}
\begin{aligned}
\label{DefAuxiliaryFctH}
  &\mathcal{H}(E)=(E+p_0)\ln n_{\rm F}^-(E)\\
  &-T\,{\rm Li}_2\left(-\frac{1}{n_{\rm B}(E+p_0)}\right)-T\,{\rm Li}_2\left(1-\frac{1}{n_{\rm F}^-(E)}\right),
\end{aligned}
\end{equation}
one obtains:
\begin{equation}
\begin{aligned}
\label{FinalAnalytResultImSigma34}
  \Im\Sigma_3(p,-\i p_0) &= \frac{2p^2\!+\!m_{\rm M}^2}{2p^2}\,\Im\Sigma_0\!+\!\frac{\gMqSq p_0T}{16\pi p^3} \left.\mathcal{H}(E)\right|_{E_{\rm min}}^{E_{\rm max}},\\
  \Im\Sigma_4(p,-\i p_0) &= \Im\,\Sigma_0+\frac{\gMqSq T}{16\pi pp_0} \left.\,\mathcal{H}(E)\right|_{E_{\rm min}}^{E_{\rm max}}\;.
\end{aligned}
\end{equation}
These results for $\Im\,\Sigma_j(p,-\i p_0)$, $j=0,3,4$, will be used for the evaluation of the shear viscosity \eqref{GeneralDiracEtaCombinedFULL}, where $p_0=\sqrt{m^2+p^2}$ is a function of the quark momentum due to the on-shell treatment. Their dependence on the momentum $p$ is shown in Fig.~\ref{Fig:ImaginaryPartsOfSigmas} for the pion case at two different temperatures, $T=220,260\;{\rm MeV}$, and vanishing chemical potential. Due to the explicit analytical form of the self-energy contributions, they can be easily implemented and numerical issues arise only from handling the peaks in the integrand of the Kubo formula \eqref{GeneralDiracEtaCombinedFULL}, cf. also Ref.~\cite{LangWeise:2014}. We emphasize that its energy and momentum integrals are carried out independently and the functions $A,B,C,D$ in its integrand remain off-shell.

\subsection{Off-shell quarks} 
So far we have treated the external quark in the Fock self-energy $\Sigma_\beta^{\rm S/P}(p,-\i p_0)$ in Eq.~\eqref{FockDiagramSigmaBetaSP} as an on-shell particle with $p_0^2=m^2+p^2$ when determining the imaginary parts of $\Sigma_j(p,-\i p_0)$, $j=0,3,4$. According to the general Kubo formula for the shear viscosity derived in Eq.~\eqref{EtaLeadingNcAsTwoSpectralDensities}, the quark spectral function $\rho(p_0,\vec{p})$ enters for off-shell kinematics. Using on-shell expressions for $\Im\,\Sigma_j$ is a commonly used but unnecessary approximation. While in the on-shell approximation only one dissipative process (meson decay into a quark-antiquark pair) gives rise to an imaginary part, the off-shell situation features several dissipative processes. In Appendix \ref{AppOffShell} the detailed analytical calculation of the off-shell imaginary parts of the quark self-energy from mesonic fluctuations is presented. Here we state only the results:

\begin{equation}
\begin{aligned}
\label{OffShellIm04Reminder}
  \Im\,\Sigma_0^{\rm off} &= \frac{g_{\rm \pi qq}^2}{16\pi p}\left(J^{\rm I}+J^{\rm II}+J^{\rm III}\right)\;, \\
  \Im\,\Sigma_4^{\rm off} &= \frac{g_{\rm \pi qq}^2}{16\pi pp_0}\left(K^{\rm I}+K^{\rm II}+K^{\rm III}\right)\;,
\end{aligned}
\end{equation}
and
\begin{equation}
\label{OffShellIm3}
  \Im\,\Sigma_3^{\rm off}=\frac{m_{\rm M}^2+p^2-p_0^2-m^2}{2p^2}\,\Im\,\Sigma_0^{\rm off}+\frac{p_0^2}{p^2}\,\Im\,\Sigma_4^{\rm off}\;.
\end{equation}
It is interesting to note that in contrast to the on-shell results in Eqs.~\eqref{FinalAnalytResultImSigma0} and \eqref{FinalAnalytResultImSigma34}, the off-shell imaginary parts feature a vacuum contribution included in $J^{\rm III}$ and $K^{\rm III}$ which do not vanish in the limit $T,\mu\to 0$.

\section{Results for the shear viscosity} \label{Results}
\begin{figure}[t!]
\begin{center}
  \includegraphics[width=0.49\textwidth]{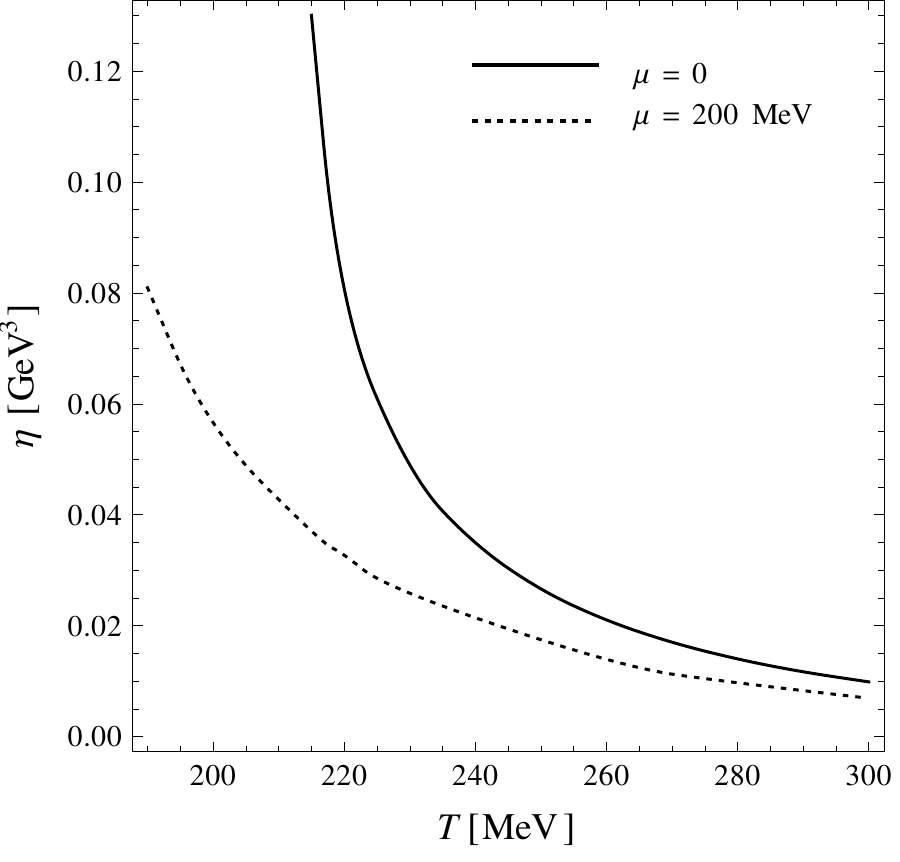}
  \caption{Temperature dependence of shear viscosity calculated from the NJL model in its large-$\Nc$ expansion for vanishing quark chemical potential and $\mu=200\;{\rm MeV}$. See the discussion in the text.}
  \label{Fig:ResultsEtasOnShell}
\end{center}
\end{figure}

We are now ready to present results for the shear viscosity $\eta$ in the NJL model. First we use the on-shell expressions for the imaginary parts $\Im\,\Sigma_j$, $j=0,3,4$, written in Eqs.~\eqref{FinalAnalytResultImSigma0} and \eqref{FinalAnalytResultImSigma34} to evaluate the Kubo formula \eqref{GeneralDiracEtaCombinedFULL} numerically. The temperature dependence of the viscosity is shown in Fig.~\ref{Fig:ResultsEtasOnShell} for two values, $\mu=0,200\;{\rm MeV}$, of the quark chemical potential. Due to the on-shell restrictions only the temperature range above the Mott temperature, $T>T_{\rm M}$, is accessible. One has $T_{\rm M}(\mu=0)=212\;{\rm MeV}$ and $T_{\rm M}(\mu=200\;{\rm MeV})=171\;{\rm MeV}$. We observe an overall decreasing function $\eta(T)$ and also decreasing values $\eta(\mu)$ for increasing the chemical potential.
A small shear viscosity reflects a highly correlated system: stronger interactions with the thermal medium lead to a lower value of $\eta$ \cite{LangWeise:2014}. We conclude that the quark plasma described by the NJL model, where the shear viscosity is induced by mesonic fluctuations occurring at order $1/\Nc$, becomes more strongly correlated for both increasing temperature and chemical potential.

\begin{figure}[t!]
\begin{center}
  \includegraphics[width=0.49\textwidth]{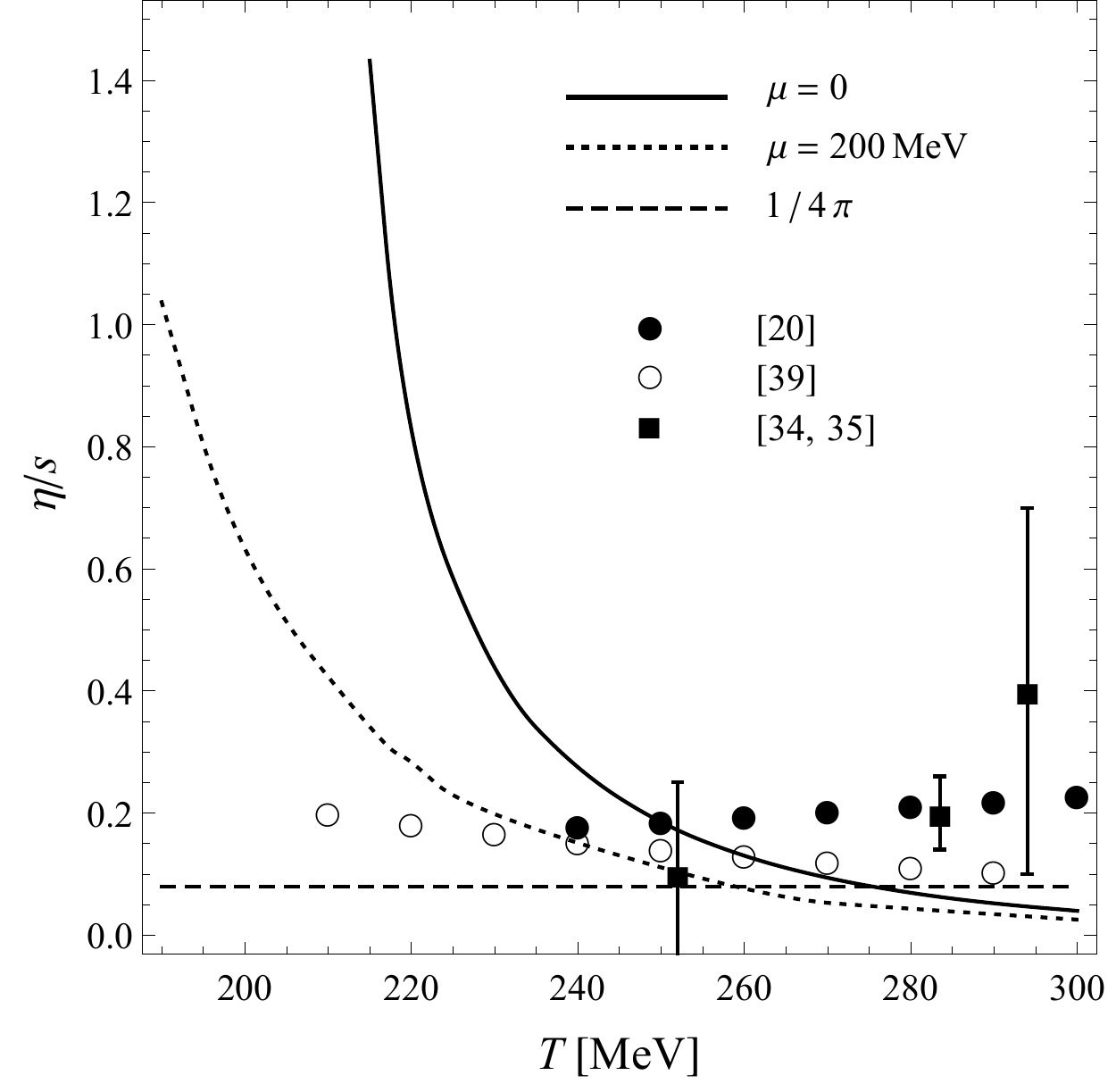}
  \caption{Temperature dependence of the ratio $\eta/s$ for vanishing quark chemical potential and $\mu=200\;{\rm MeV}$. See the discussion in the text.}
  \label{Fig:ResultsEtasOnShellRatio}
\end{center}
\end{figure}

Now we turn to the ratio $\eta/s$, shear viscosity to entropy density. Consistent with the $1/\Nc$-approach, we use for $s$ the entropy density of non-interacting constituent quarks with $(T,\mu)$-dependent masses:
\begin{equation}
\label{OurEntropyDensity}
\begin{aligned}
  s(T,\mu) &=\frac{\Nc\Nf}{\pi^2}\int_0^{\Lambda,\infty}\dint p\,p^2\,
 \left[-\ln n_{\rm F}^+(E) - \ln n_{\rm F}^-(E) \right. \\
  & \left. \;\;\;\;\; + \beta(E+\mu)n_{\rm F}^+(E)+\beta(E-\mu)n_{\rm F}^-(E)\right],
\end{aligned}
\end{equation}
where the upper boundary of the momentum integral, $\int^{\Lambda,\infty}$, encodes $E=\sqrt{m^2+p^2}$ for $p<\Lambda$ but $E=\sqrt{m_0^2+p^2}$ for $p>\Lambda$, with the current-quark mass $m_0$ instead of the constituent-quark mass $m$. This so-called soft-cutoff scheme ensures the correct Stefan-Boltzmann limit of $s(T,\mu)$ at high temperatures. Inspection of Fig.~\ref{Fig:ResultsEtasOnShellRatio} shows that the overall scale of the ratio $\eta/s$ is comparable to $1/4\pi$ \cite{Maldacena99,Kovtun05}. However, for large enough temperatures it undershoots the AdS/CFT benchmark as it can happen also in other quantum field theoretical models \cite{Cohen07,Rebhan12,Mamo12}. In the NJL model this happens for vanishing chemical potential at $T\approx 275\;{\rm MeV}$, and for a finite chemical potential, $\mu=200\;{\rm MeV}$, at a somewhat lower temperature $T\approx 260\;{\rm MeV}$. 

Furthermore, we compare our results for $\eta/s$ to those from lattice QCD, \cite{Nakamura2005,Meyer2007}, which are shown as squares with error bars in Fig.~\ref{Fig:ResultsEtasOnShellRatio}. They have been obtained within pure-gauge QCD and suggest a rising ratio $\eta/s$ for $T>250\;{\rm MeV}$, a behavior which is not found in the NJL model. This qualitative difference can be explained by considering results from hard thermal loop (HTL) calculation in QCD \cite{Arnold2000,Arnold2003}. At leading logarithmic order one finds the behavior \cite{Kapusta2006}:
\begin{equation}
\label{EtaKapusta}
  \eta=\frac{C_1 T^3}{\alpha_{\rm s}^2\,\ln (C_2/\alpha_{\rm s})}\;,
\end{equation}
with flavor-dependent coefficients $C_1$ and $C_2$. Consequently, the dimensionless ratio $\eta/s$ scales as $\eta/s\sim \alpha_{\rm s}^{-2}$ at leading order. For increasing temperature the QCD coupling becomes weak, $\alpha_{\rm s}\to 0$, and the ratio $\eta/s$ rises with $T$ according to the HTL results. Lattice QCD suggests that this trend sets in at rather low temperatures, where HTL calculations are not applicable since they are based on perturbative-QCD and resummation techniques. The main reason for the rising behavior of $\eta/s$ in lattice QCD are the weaker correlations between the gauge bosons towards asymptotic freedom. In contrast to this, the NJL-model coupling $G$ remains constant and the viscous effects from mesonic fluctuations are growing in the considered temperature range \mbox{$180\;{\rm MeV}\lesssim T\lesssim 300\;{\rm MeV}$}.
As a consequence, the NJL model provides $\eta$ and $\eta/s$ decreasing with $T$ and $\mu$. Note that for large $T$ the thermal quark mass, $m(T)\sim gT$, dominates the constituent-quark mass derived within the NJL model. In our results the quark mass becomes small in this temperature region: $m \to m_0$, cf. Fig.~\ref{Fig:ThermalMassesStandard}.

The open circles in Fig.~\ref{Fig:ResultsEtasOnShellRatio} are the results from Ref.~\cite{Plumari2012}, where the shear viscosity has been evaluated from the basic Kubo formula~\eqref{KuboEta} using cross sections $\sigma_{\rm tot}$ from a parton cascade model with elastic two-body collisions for gluons only. Their results are described by
\begin{equation}
  \label{KuboResultsPlumari}
\frac{\eta}{s}=\frac{0.195}{\sigma_{\rm tot}T^2}\;,
\end{equation}
and $\sigma_{\rm tot}=9\,{\rm mb}=0.9\;{\rm fm}^2$ has been used to obtain the open circles in Fig.~\ref{Fig:ResultsEtasOnShellRatio}. In comparison to our NJL results one gets a decreasing but flatter ratio $\eta/s$ in that approach. The assumption of a temperature-independent total cross section $\sigma_{\rm tot}$ does not describe the high-$T$ behavior of HTL calculations and suggested by lattice QCD.

\begin{figure}[t!]
\begin{center}
  \includegraphics[width=0.49\textwidth]{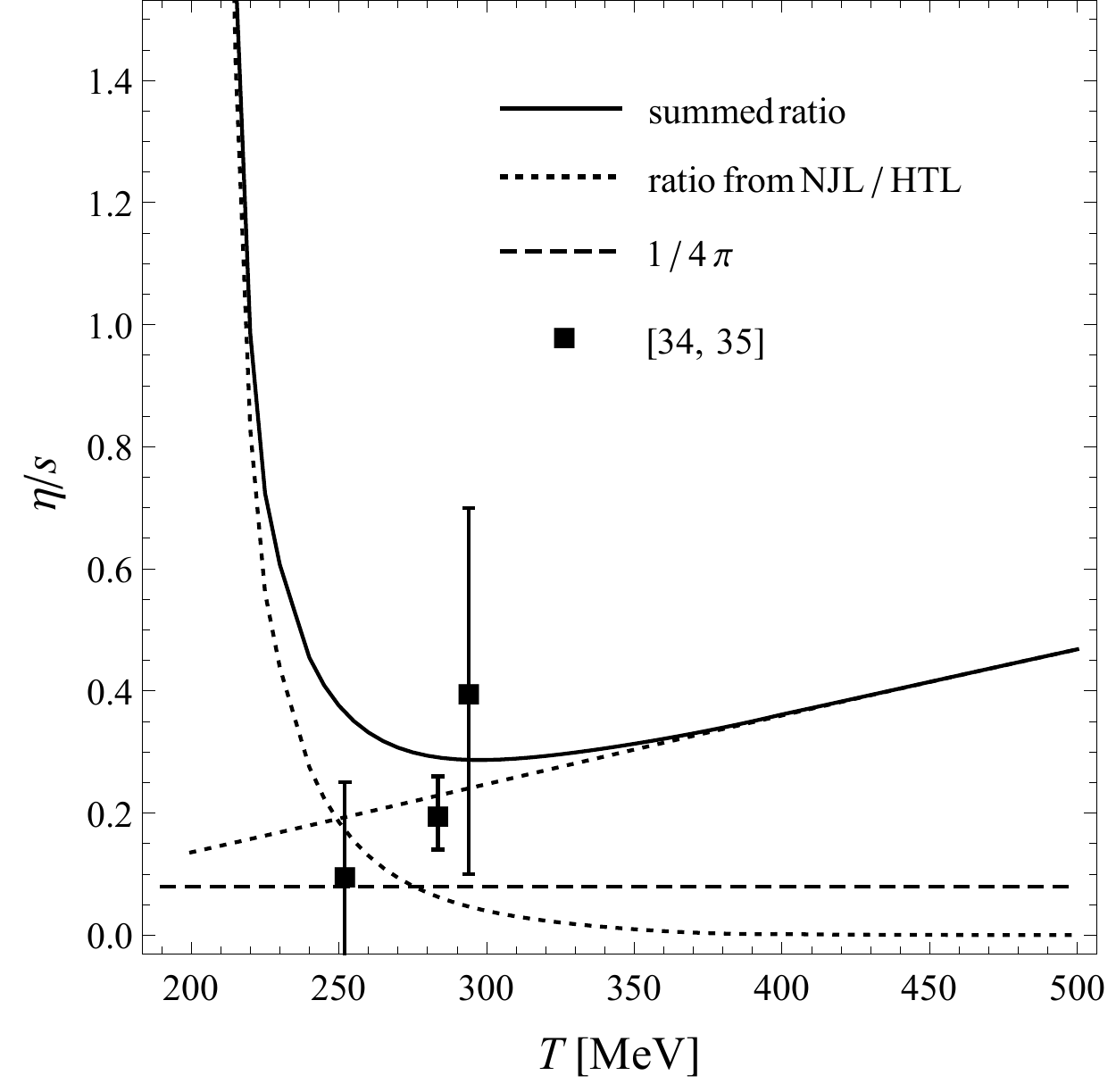}
  \caption{Individual and summed ratios $\eta/s(T)$ from the NJL model (low-$T$ region) and from HTL calculations (high-$T$ region) at vanishing quark chemical potential. See the discussion in the text.}
  \label{Fig:ResultsEtasOnShellInclHTL}
\end{center}
\end{figure}

The rising behavior of $\eta/s$ has been parameterized in Ref.~\cite{Christiansen2014} as
\begin{equation}
  \left.\frac{\eta}{s}\right|_{\rm HTL}=\frac{a}{\alpha_s^\gamma}\;,
\end{equation}
with $a=0.2$ and $\gamma=1.6$ extracted from a combined fit to results from functional-renormalization-group methods and HTL calculations. Note that $\gamma\approx 2$ as expected from the pure HTL result for gauge theories, Eq.~\eqref{EtaKapusta}. Ref.~\cite{Christiansen2014} has used the following form for the temperature dependence of the running QCD coupling \cite{Nesterenko2000,Nesterenko2001,Nesterenko2003}:
\begin{equation}
  \alpha_{\rm s}(T)=\frac{4\pi}{\beta_0}\frac{z^2-1}{z^2\,\ln z^2}\;,
\end{equation}
with the beta-function coefficient $\beta_0=(11\Nc-2\Nf)/3$, and the reduced temperature $z=0.79\,T/T_{\rm c}$, where $T_{\rm c}\approx 155\;{\rm MeV}$.

The shear viscosity $\eta$ obtained from HTL calculations is induced by dissipative processes in the gauge sector, whereas $\eta$ calculated from the NJL model arises from mesonic fluctuations in the quark sector. We interpolate between the low-$T$ (NJL) and high-$T$ (HTL) domains by taking the sum of the two corresponding ratios $\eta/s$, as it is suggested in Ref.~\cite{Christiansen2014}. The resulting summed ratio is shown as the solid line in Fig.~\ref{Fig:ResultsEtasOnShellInclHTL}. It develops a minimum at $T_{\rm min}=295\;{\rm MeV}$ with $\eta/s(T_{\rm min})=0.29\gtrsim 3.6/4\pi$ due to the change between quarks and gluons as active degrees of freedom. In comparison to the analogous results $T_{\rm min}^{\rm QCD}=200\;{\rm MeV}$ and $\eta/s(T_{\rm min}^{\rm QCD})=0.17$ from Ref.~\cite{Christiansen2014}, both the minimal value of $\eta/s$ and its location are shifted to higher values in the present work. The main reason for this is the rather high chiral crossover temperature, $T_{\rm c}=190\;{\rm MeV}
$, in the two-flavor NJL model. Clearly, Figure~\ref{Fig:ResultsEtasOnShellInclHTL} should be taken just for qualitative orientation.
The position and value of the minimum in $\eta/s$ depend sensitively on the detailed interpolation between the low-$T$ and high-$T$ domains. Taking the sum of the corresponding ratios at low and high temperature is only one possible way.

\begin{figure}[t!]
\begin{center}
  \includegraphics[width=0.49\textwidth]{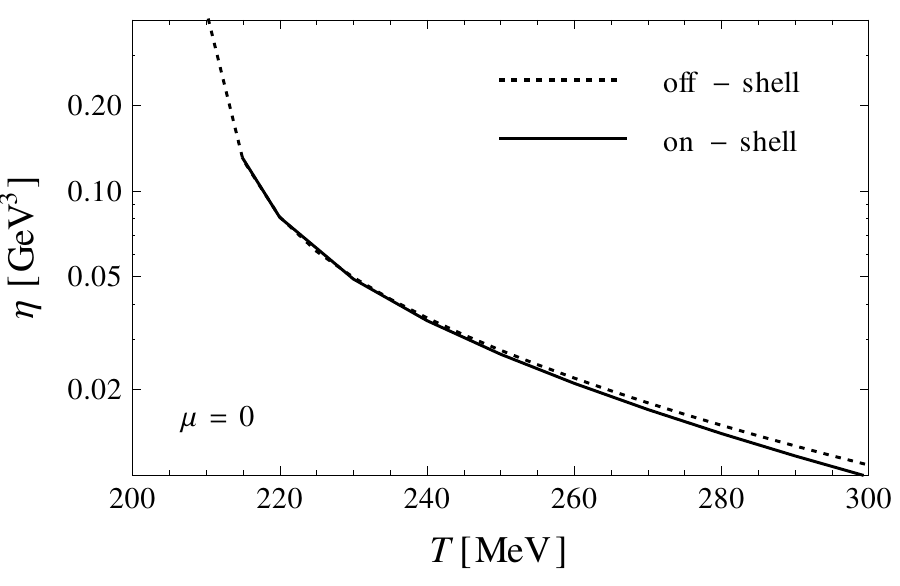}
  \includegraphics[width=0.49\textwidth]{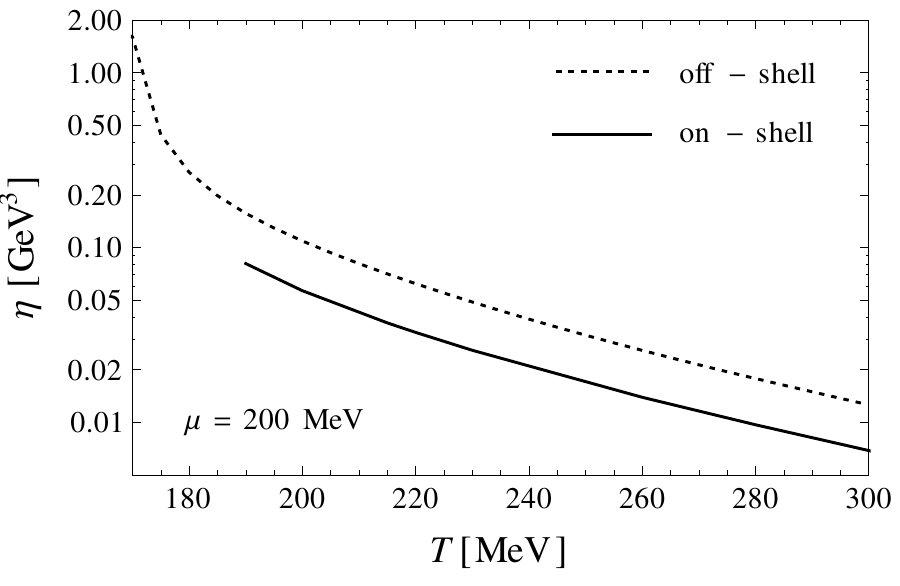}
  \caption{Comparison between the on-shell and off-shell calculation of the shear viscosity at vanishing quark chemical potential (left panel) and $\mu=200\;{\rm MeV}$ (right panel)}
  \label{Fig:ResultsEtasOffShell}
\end{center}
\end{figure}

We have also numerically evaluated the Kubo formula for the shear viscosity \eqref{GeneralDiracEtaCombinedFULL} using results for the off-shell spectral functions $\rho_j=\Im\,\Sigma_j$, $j=0,3,4$ given in Eqs.~\eqref{OffShellIm04Reminder} and \eqref{OffShellIm3}. The corresponding results are shown in Fig.~\ref{Fig:ResultsEtasOffShell} and compared. The new feature of the proper off-shell treatment is the presence of viscous processes in the whole temperature region. Therefore, the constituent-quark mass does not provide any restriction on a finite shear viscosity. It is interesting to observe that at $\mu=0$ the results at small $T$ smoothly join those of the on-shell approximation where the Mott-condition $m_{\rm M}>2m$ has to be fulfilled. The quantitative difference in this region is almost negligible. At $\mu=200\;{\rm MeV}$ the difference is an almost constant factor shifting the viscosity to higher values, but the overall behavior of $\eta(T,\mu)$ is not changed.

One can explain this qualitative agreement by the peaking of the integrand in the Kubo formula: the main contribution to $\eta$ in Eq.~\eqref{GeneralDiracEtaCombinedFULL} is collected around the minimum of the denominator $D(\epsilon,p)$, cf. Fig.~2 in Ref.~\cite{LangWeise:2014}. This essentially leads to the on-shell approximation as one can argue with Eqs.~\eqref{DefAuxiliaryFctN1N2} and \eqref{ResultsABCD}:
\begin{equation}
  D\to 0 \;\;\;\; \Leftrightarrow \;\;\;\; N_1, N_2 \to 0 \;\;\;\; \Rightarrow \;\;\;\; p_0^2-\vec{p}^2-m^2=0\;.
\end{equation}
We conclude that the off-shell treatment provides only subleading corrections due to the peak structure of the integrand of the Kubo formula. One would expect that the shear viscosity becomes smaller in the off-shell treatment since more dissipative processes are at work. But on the contrary, the complicated arrangement of the imaginary parts $\rho_j$, $j=0,3,4$, in the integrand of $\eta$ leads eventually to an increasing shear viscosity compared to the on-shell approximation.

\section{Summary and Conclusion} \label{Summary}
In this work we have investigated the shear viscosity of hot and dense quark matter described by a large-$\Nc$ NJL model for two flavors. We have used the Kubo formalism and calculated the shear viscosity $\eta$ from a thermal quark spectral function with inclusion of its full Dirac structure. Instead of a single width there are now three (off-shell) imaginary parts which determine the positive-definite shear viscosity.

In the large-$\Nc$ counting, the dominant dissipative process arises from mesonic fluctuations. They are dynamically generated by virtual quark-antiquark loops resummed to all orders in the non-perturbative Bethe-Salpeter equation. The mesonic Fock contribution to the gap equation are of subleading order $1/\Nc$. We have calculated the three components $\hat{\Sigma}_j$ of the quark self-energy provided by the mesonic Fock term, both for on-shell and off-shell kinematics. Evaluating the Kubo formula with the input $\rho_j=\Im\;\hat{\Sigma}_j$, we have found a decreasing shear viscosity as function of both temperature and quark chemical potential. At vanishing chemical potential, the proper off-shell treatment extends the on-shell approximation into the low-temperature region where the on-shell viscous effects are kinematically excluded.
Apart from this, off-shell corrections have no further quantitative or qualitative influence. However, at finite quark chemical potentials, off-shell effects shift the shear viscosity to higher values but its overall qualitative behavior is not changed.

We have observed that the dimensionless ratio $\eta/s$ undershoots the AdS/CFT benchmark $1/4\pi$ at large enough temperatures in the NJL model. Combining our results for the shear viscosity with perturbative results from hard-thermal-loop calculations in the high-$T$ region, we find that the ratio $\eta/s$ develops a minimum well above the AdS/CFT benchmark. The interpolated results compare reasonably with those from lattice QCD regarding the overall behavior and scale of the ratio $\eta/s$. However, since the chiral crossover temperature in the two-flavor NJL model, $T_{\rm c}\approx 190\;{\rm MeV}$, is larger than the lattice-QCD result, $T_{\rm c}\approx 155\;{\rm MeV}$, the onset of the dominant viscous effects is shifted to higher temperatures by the Mott condition.

In summary one can conclude that the correlated quark matter described by the NJL model features a small ratio shear viscosity over entropy density as it is characteristic for a perfect fluid.

\section*{Acknowledgments}
This work is partially supported by BMBF and by the DFG Cluster of Excellence ``Origin and Structure of the Universe''. Useful discussions with T. Hatsuda and Y. Hidaka are gratefully acknowledged. R. Lang thanks the ECT* Trento for kind hospitality. He has been supported also by the TUM Graduate School (TUM-GS) and by the RIKEN IPA and iTHES projects.

\appendix
\section{Meson propagators from the BSE} \label{AppBSE}
The meson masses are derived from the Bethe-Salpeter equation (BSE) as pole-masses of resummed quark-antiquark scattering modes:
\begin{equation}
\label{RPAresummedAPP}
  D_{\rm M}=G+G\Pi^{\rm S/P} D_{\rm M}=\frac{G}{1-\Pi^{\rm S/P} G}\;,
\end{equation}
where $\Pi^{\rm S/P}$ denotes the polarization tensor:
\begin{equation}
\begin{aligned}
\label{GenStructQuAQuLoopCalStep1}
  &\Pi^{\rm S/P}(\vec{p},\omega_n)\\ 
  &\!\!=8\Nc\ThermalInt{m}{q}\frac{\mp m^2+\nu_m(\nu_m-\omega_n)+\vec{q}(\vec{q}-\vec{p})}{[\nu_m^2+E_q^2][(\nu_m-\omega_n)^2+E_\Delta^2]}\\
  &\!\!=4\Nc\ThermalInt{m}{q}\left[\frac{1}{\nu_m^2+E_q^2}+\frac{1}{(\nu_m-\omega_n)^2+E_\Delta^2}\right]\\
  &\!\!+4\Nc N^{\rm S/P}\ThermalInt{m}{q}\frac{1}{[\nu_m^2+E_q^2][(\nu_m-\omega_n)^2+E_\Delta^2]}\;,
\end{aligned}
\end{equation}
with the energies $E_q^2=\vec{q}^2+m^2$ and $E_\Delta^2=(\vec{q}-\vec{p})^2+m^2$. It can be expressed as
\begin{equation}
\label{QuarkAntiquarlLoopResultAPP}
  \Pi^{\rm S/P}(\vec{p},\omega_n)=8\Nc I_1+2\Nc N^{\rm S/P} I_2(\vec{p},\omega_n)\;,
\end{equation}
where we have denoted the momentum-independent part of the tensor by $I_1$, whereas the momentum dependence is encoded in $I_2(\vec{p},\omega_n)$. In addition we have introduced $N^{\rm P}=-\left(\omega_n^2+\vec{p}^2\right)$ and $N^{\rm S}=N^{\rm P}-4m^2$ describing the pion and sigma modes, respectively. We have defined:
\begin{widetext}
\begin{equation}
\begin{aligned}
\label{DefIntegralI2}
  I_2(\vec{p},\omega_n)& =\ThermalInt{m}{q}\frac{1}{[\nu_m^2+E_q^2][(\omega_n-\nu_m)^2+E_\Delta^2]}\\
  &=\int\frac{\dint^3q}{(2\pi)^3}\;\frac{1}{4E_qE_\Delta}\left[\frac{2E_+}{\omega_n^2+E_+^2}\right. +\;\frac{(\i\omega_n-E_+)\left(n_{\rm F}^-(E_q)+n_{\rm F}^+(E_\Delta)\right)-(\i\omega_n+E_+)\left(n_{\rm F}^+(E_q)+n_{\rm F}^-(E_\Delta)\right)}{\omega_n^2+E_+^2}\\
  & \hspace{4.4cm} \left. +\;\frac{(\i\omega_n+E_-)\left(n_{\rm F}^+(E_q)-n_{\rm F}^+(E_\Delta)\right)-(\i\omega_n-E_-)\left(n_{\rm F}^-(E_q)-n_{\rm F}^-(E_\Delta)\right)}{\omega_n^2+E_-^2} \right],
\end{aligned}
\end{equation}
\end{widetext}
having defined $E_\pm=E_q\pm E_\Delta$. The momentum-independent part of the polarization tensor reads

\begin{equation}
\begin{aligned}
\label{DefIntegralI1RelGapPiSP}
  I_1&=\ThermalInt{m}{q}\frac{1}{\nu_m^2+q^2+m^2}\\ 
  &=\frac{1}{4\pi^2}\int_0^\Lambda\dint p\frac{p^2}{E_p}\,\big(1-n_{\rm F}^+(E_p)-n_{\rm F}^-(E_p)\big) \;.
\end{aligned}
\end{equation}
As mentioned already before, $I_1$ refers also to the chiral condensate, which can be seen from inspecting Eqs.~\eqref{GapEquationHartree} and \eqref{ThermalGapEquation}:

\begin{equation}
  I_1=\frac{m-m_0}{8G\Nc m}=-\frac{\langle\psib\psi\rangle}{8\Nc m}\;.
\end{equation}
In the calculation of $\Pi^{\rm S/P}$ one has to take care of several minus signs: the fermion loop gives a global minus sign, in the pseudoscalar channel one has $\i^2=-1$ and additionally $\{\gamma_5,\gamma_\mu\}=0$ and $\{\gamma_i,\gamma_j\}=-2\delta_{ij}$ for the Euclidean gamma matrices.

\section{Details of the on-shell calculation} \label{AppOnShell}
The individual contributions are given by
\begin{equation}
  \label{ThermalSigma034NoMatsubaraSum}
\begin{aligned}
  \Sigma_0 &= \gMqSq\ThermalInt{m}{q} \frac{1}{\nu_m^2+E_f^2}\frac{1}{(\nu_n-\nu_m)^2+E_b^2}\;, \\
  \Sigma_3 &= \gMqSq\ThermalInt{m}{q} \frac{\vec{p}\cdot\vec{q}}{p^2}\frac{1}{\nu_m^2+E_f^2}\frac{1}{(\nu_n-\nu_m)^2+E_b^2}\;,\\
  \Sigma_4 &= \gMqSq\ThermalInt{m}{q} \frac{\nu_m}{\nu_n}\frac{1}{\nu_m^2+E_f^2}\frac{1}{(\nu_n-\nu_m)^2+E_b^2}\;,
\end{aligned}
\end{equation}
with the energies $E_f^2=\vec{q}^2+m^2$ and $E_b^2=(\vec{p}-\vec{q})^2+m_{\rm M}^2$. As always, the Matsubara sums can be carried out leading to some finite result with a combination of Bose and Fermi distribution functions. We arrive at
\begin{equation}
\label{ThermalSigma03AfterMatsubara}
\begin{aligned}
  \Sigma_{0,3}(p,\nu_n) &=\gMqSq \int\frac{\dint^3q}{(2\pi)^3}\,\mathcal{F}_{0,3}\left[\frac{1}{2E_bE_f}\right.\\
  &\left. \hspace{-2cm}\times\left(\frac{E_+ Z_1}{E_+^2+\nu_n^2}+\frac{E_- Z_2}{E_-^2+\nu_n^2}\right)+\frac{\i\nu_n\,Z_3}{(E_+^2+\nu_n^2)(E_-^2+\nu_n^2)}\right],
\end{aligned}
\end{equation}
with $E_\pm=E_b\pm E_f$. The quark-meson coupling, $g_{\rm Mqq}$, can be pulled out of the integral since no momentum dependence is taken into account as it has been justified before. We have introduced $\mathcal{F}_{0,3}$ as
\begin{equation}
\begin{aligned}
\label{DefinitionF03}
  \mathcal{F}_0 &= 1\;,\\
  \mathcal{F}_3 &= \frac{\vec{p}\cdot\vec{q}}{p^2}=\frac{m_{\rm M}^2-2m^2-2E_fp_0}{2p^2}\;,
\end{aligned}
\end{equation}
and have denoted the combinations of Bose and Fermi distributions as $Z_i(E_b,E_f)$:
\begin{equation}
\begin{aligned}
  Z_1 &=1+n_{\rm B}(E_b)-\frac 12\left(n_{\rm F}^+(E_f)+n_{\rm F}^-(E_f)\right), \\
  Z_2 &= n_{\rm B}(E_b)+\frac 12\left(n_{\rm F}^+(E_f)+n_{\rm F}^-(E_f)\right)>0\;, \\ 
  Z_3 &=n_{\rm F}^+(E_f)-n_{\rm F}^-(E_f)>0\;.
\end{aligned}
\end{equation}
The Bose and Fermi distributions read
\begin{equation}
\label{DefBoseFermiMuPlusMinusDistr}
  n_{\rm B}(E)=\frac{1}{\e^{\beta E}-1}\;, \;\;\; n_{\rm F}^\pm(E)=n_{\rm F}(E\mp\mu)=\frac{1}{\e^{\beta (E\mp\mu)}+1}\;,
\end{equation}
where the $\pm$ signs denote quark and antiquark distribution functions, respectively. When carrying out the Matsubara sum also for the $\Sigma_4$ part of the self-energy, we get:
\begin{equation}
\begin{aligned}
\label{ThermalSigma4AfterMatsubara}
  \Sigma_4(p,\nu_n)&=\gMqSq \int\frac{\dint^3q}{(2\pi)^3}\,\left[\frac{1}{2E_b}\right.\\
  &\left. \hspace{-1.6cm}\times\left(\frac{Z_1}{E_+^2+\nu_n^2}+\frac{Z_2}{E_-^2+\nu_n^2}\right)-\frac{(E_b^2-E_f^2+\nu_n^2)Z_3}{2\i\nu_n(E_+^2+\nu_n^2)(E_-^2+\nu_n^2)}\right].
\end{aligned}
\end{equation}
The non-vanishing imaginary parts of $\Sigma_i$ are induced by their pole structure:
\begin{equation}
\begin{aligned}
\label{DeltaFctArrImPart}
  \lim_{\eps\to 0}\Im\left.\frac{Z}{x^2+\nu_n^2}\right|_{\nu_n\mapsto -\i p_0+\eps} &= Z\pi\delta(x^2-p_0^2)\\
  &\hspace{-2cm}=\frac{\pi Z}{2p_0}\left(\delta(x-p_0)+\delta(x+p_0)\right).
\end{aligned}
\end{equation}
This means for the $Z_1$ term: $E_f+E_b\pm p_0=0$, where only the minus sign can be realized. For the $Z_2$ term, $E_f-E_b\pm p_0=0$, both signs can be realized for the time being. We will see that only the plus-sign case contributes to the (on-shell) imaginary parts, so there is just one contribution from $Z_2$. Later, the $Z_3$ term is considered separately. We start with the first two terms $Z\in\{Z_1,Z_2\}$. Using the identify \eqref{DeltaFctArrImPart} we find the following structure when evaluating $\Sigma_{0,3}(\vec{p},\nu_n)$ from Eq.~\eqref{ThermalSigma03AfterMatsubara} after analytical continuation has been carried out:
\begin{equation}
\label{CalcZ1Z2PhaseSpace}
\begin{aligned}
  &\int\frac{\dint^3q}{(2\pi)^3}\frac{\pi Z}{2p_0}\frac{1}{2E_bE_f}\,\delta(E_b-(*))\\
  &=\int\frac{\dint^3q}{(2\pi)^3}\frac{\pi Z}{2p_0E_f}\,\delta(E_b^2-(*)^2)\\
  &=2\pi\int_{-1}^1\dint\xi\int_0^\infty \frac{\dint q\, q^2}{(2\pi)^3}\frac{\pi Z}{2p_0E_f }\delta(E_b^2(\xi)-(*)^2)\\
  &=2\pi\int_{m}^\infty\frac{\dint E_f}{(2\pi)^3}\frac{\pi Z}{4p_0p}\,\Theta(1-\xi^2)\;,
\end{aligned}
\end{equation}
where $\xi=\cos\theta$. In order to carry out the integral over the delta function we have used
\begin{equation}
\begin{aligned}
\label{EbInTermsOfXi}
  E_b^2 &= m_{\rm M}^2+(\vec{p}-\vec{q})^2=m_{\rm M}^2+p^2+q^2-2pq\xi\;, \\
  &\Rightarrow \left|\frac{\partial E_b^2}{\partial\xi}\right|=2pq\;,
\end{aligned}
\end{equation}
and converted the momentum integral to an energy integral using $q\,\dint q=E_f\,\dint E_f$. The ill-conditioned $\Theta$ term can be removed by the following consideration: from Eq.~\eqref{EbInTermsOfXi} it is clear that $|\xi|\leq 1$ is fulfilled if and only if
\begin{equation}
\label{DefCapitalFfunctionPhaseSpace}
  -1 \leq \frac{E_b^2-m_{\rm M}^2-p^2-q^2}{2pq}\leq 1\;,
\end{equation}
equivalent to $F(E_f,p)\geq 0$, where we have defined
\begin{equation}
\begin{aligned}
  F(E_f,p)&=4p^2 (E_f^2-m^2)\\
  &-\left[E_b^2-m_{\rm M}^2-p^2+m^2-E_f^2\right]^2\;.
\end{aligned}
\end{equation}
For a given value of the absolute momentum the roots of $F(\,\cdot\, ,p)$ read for the plus-sign case $0=E_f+E_b+p_0$, and therefore $E_b^2=(E_f+p_0)^2$:
\begin{equation}
\label{Eminmax}
\begin{aligned}
  E_{{\rm max},{\rm min}}&=\frac{1}{2m^2}\left[\left(m_{\rm M}^2-2m^2\right)\sqrt{m^2+p^2}\right.\\
  &\hspace{1.5cm}\left.\pm p\,m_{\rm M}\sqrt{m_{\rm M}^2-4m^2}\right] \\
\end{aligned}
\end{equation}
The range of integration, $E_f\in[E_{\rm min},E_{\rm max}]$, depends therefore linearly on the external quark momentum:
\begin{equation}
\label{EmaxEminDifference}
  E_{{\rm max}}-E_{{\rm min}} =\frac{p\,m_{\rm M}}{m^2}\sqrt{m_{\rm M}^2-4m^2}\;.
\end{equation}
In the limit of a vanishing external quark momentum the range of integration collapses to one single point:
\begin{equation}
\label{EmaxEminZeroValue}
  \left.E_{{\rm max},{\rm min}}\right|_{p=0} =\frac{m_{\rm M}^2}{2m}-m>m\;.
\end{equation}
We emphasize that the whole discussion is only valid for temperatures above the Mott temperatures $T_{\rm M}$, where the pion mass is at least twice the constituent-quark mass. This constraint can be seen explicitly from the integral boundaries Eq.~\eqref{Eminmax}. We have already introduced the Mott temperature when discussing thermal quark and meson masses, where $T_{\rm M}\approx 212\;{\rm MeV}$ have been found in the case of vanishing quark chemical potential. Note that this discussion remains valid also in the chiral limit, where the current-quark mass is set to zero, $m_0=0$. In this case, the pion mass vanishes in the Nambu-Goldstone phase at low temperatures but it is finite when chiral symmetry is restored for large temperatures.

We conclude that under the condition $m_{\rm M}>2m$, i.e. for $T>T_{\rm M}$, the phase space is always non-empty and compact: $\emptyset\neq [E_{\rm min},E_{\rm max}]\subseteq [m,\infty)$. This fact implies that the shear viscosity $\eta$ will evaluate to some finite result in this temperature region. However, we have also derived the following substitution rule
\begin{equation}
  \int_{m}^\infty\frac{\dint E_f}{(2\pi)^3} \big(\cdot\big) \Theta(1-\xi^2) = \int_{E_{\rm min}}^{E_{\rm max}} \frac{\dint E_f}{(2\pi)^3} \big(\cdot\big)\;,
\end{equation}
which leads finally to a well-conditioned one-dimensional numerical integral.

For the sake of completeness, we also mention the minus-sign case, i.e. $0=E_b+E_f-p_0$. If we plug in $E_b^2=(E_f-p_0)^2$  into the condition \eqref{DefCapitalFfunctionPhaseSpace} then the phase space simply vanishes for any incoming quark momentum, since the range of integration would be restricted to negative energies in the fermion loop:
\begin{equation}
  E_{\rm min}'=-E_{\rm max}\, ,\;\;\; E_{\rm max}'=-E_{\rm min}\;.
\end{equation}
We can therefore conclude that only the plus-sign case, $E_b=E_f+p_0$, allows for an on-shell condition for the mesonic fluctuation. Knowing this we can now continue with the third term, $Z_3$, in Eq. \eqref{ThermalSigma03AfterMatsubara}:
\begin{equation}
\begin{aligned}
  \lim_{\eps\to 0} &\;\Im \left.\frac{\i\nu_n Z_3}{[(E_f+E_b)^2+\nu_n^2][(E_f-E_b)^2+\nu_n^2]}\right|_{\i\nu_n\mapsto p_0+\i\eps}\\
  &\;\;\;=p_0\pi Z_3\,\delta\left([(E_f+E_b)^2-p_0^2][(E_f-E_b)^2-p_0^2]\right)\\
  &\;\;\;=\frac{p_0\pi Z_3}{2p_0}\,\delta\big(\underbrace{[(E_f+E_b)^2-p_0^2]}_{=4E_fE_b}[E_b-E_f-p_0]\big)\\
  &\;\;\;=\frac{\pi Z_3}{4E_f}\,\delta\left(E_b^2-(E_f+p_0)^2\right).
\end{aligned}
\end{equation}
Note that due to the $\i\nu_n$ factor in the first line, the $p_0$ terms cancel in the final result. As done in the calculation \eqref{CalcZ1Z2PhaseSpace} the momentum integral can be performed:
\begin{equation}
\begin{aligned}
  &\int\frac{\dint^3q}{(2\pi)^3} \frac{\pi Z_3}{4E_f}\,\delta(E_b^2-(E_f+p_0)^2)\\
  &\hspace{2cm}=2\pi\int_{m}^\infty\frac{\dint E_f}{(2\pi)^3}\frac{\pi Z_3}{8p}\,\Theta(1-\xi^2)\;.
\end{aligned}
\end{equation}
Combining all contributions, we find with $\mathcal{F}_4=-\frac{E_f}{p_0}$:
\begin{equation}
\begin{aligned}
  &\Im\,\Sigma_{0,3,4}(p,-\i p_0)\\
  &=-\frac{\gMqSq}{16\pi p}\int_{E_{\rm min}}^{E_{\rm max}}\dint E_f\,\mathcal{F}_{0,3,4}\left[n_{\rm B}(E_b)+n_{\rm F}^-(E_f)\right],
\end{aligned}
\end{equation}
as it has been stated in Eq.~\eqref{SummaryOnShellImaginaryParts}.

\section{Details of the off-shell calculation} \label{AppOffShell}
Due to the symmetry properties of the quark spectral function, $\rho(-p_0,p,T,\mu)=-\rho(p_0,p,T,-\mu)$, we can simplify our discussion and restrict the off-shell energy to non-negative values $p_0\geq 0$.

We return to $\Sigma_{0,3}$ given in Eq.~\eqref{ThermalSigma03AfterMatsubara} and decompose into partial fractions for convenience:
\begin{equation}
\begin{aligned}
  \label{ThermalSigma03AfterMatsubaraPartialFraction}
\Sigma_{0,3}(\vec{p},-\i p_0) &= g_{\rm Mqq}^2\int\frac{\dint^3 q}{(2\pi)^3} \frac{\mathcal{F}_{0,3}}{4E_bE_f}\\
&\hspace{-2cm}\times\left[
\frac{1-n_{\rm F}^-(E_f)+n_{\rm B}(E_b)}{E_f+E_b+p_0}+\frac{n_{\rm B}(E_b)+n_{\rm F}^-(E_f)}{E_f-E_b+p_0+\i\eps}\right.\\
&\hspace{-2cm}\left.+\frac{n_{\rm B}(E_b)+n_{\rm F}^+(E_f)}{E_f-E_b-p_0-\i\eps}+
\frac{1+n_{\rm B}(E_b)-n_{\rm F}^+(E_f)}{E_f+E_b-p_0-\i\eps}\right].
\end{aligned}
\end{equation}
As in the on-shell discussion, taking its imaginary part probes the pole position of the partial fractions introducing four cases $\pm E_ b=E_f\pm p_0$. The fraction in the first line introduces \mbox{$E_b=-E_f-p_0<0$} which can be excluded immediately. The remaining three cases are denoted as follows:
\begin{equation}
  \label{CasesDef}
\begin{aligned}
  \mbox{Case \;\; I:} \;\; & E_b=E_f+p_0\;,\\
  \mbox{Case \, II:} \;\; & E_b=E_f-p_0\;,\\
  \mbox{Case III:} \;\; & E_b=p_0-E_f\;.
\end{aligned}
\end{equation}
As mentioned we restrict the discussion to $p_0\geq 0$. Carrying out the $\dint^3q$ integral introduces again the restriction $|\xi|\leq 1$ with $\xi=\cos\theta$ denoting the polar angle:
\begin{equation}
\label{DefCapitalFfunctionPhaseSpaceOffShell}
  -1 \leq \frac{E_b(p,p_0)^2-m_{\rm M}^2-p^2-q^2}{2pq}\leq 1\;,
\end{equation}
equivalent to $F(E_f,p,p_0)\geq 0$, where we have defined
\begin{equation}
\begin{aligned}
  F(E_f,p) &= 4p^2 (E_f^2-m^2)\\
  &-\left[E_b(p,p_0)^2-m_{\rm M}^2-p^2+m^2-E_f^2\right]^2.
\end{aligned}
\end{equation}
In the off-shell case $p>0$ and $p_0\geq  m$ are independent of each other. In the following we evaluate the three-dimensional integral \eqref{ThermalSigma03AfterMatsubaraPartialFraction} ensuring $|\xi|\leq 1$ by applying the three cases for the relation between quark and meson energy.

\begin{figure}[t!]
\begin{center}
  \subfigure[$s<0$]{\includegraphics[width=0.4\textwidth]{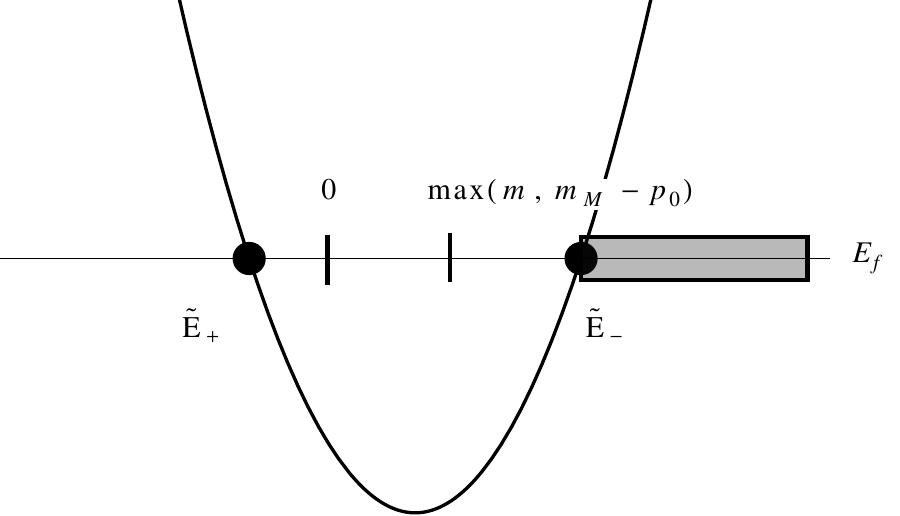}}
  \subfigure[$s>0$ with \mbox{$m<m_{\rm M}$} and $p_0<\sqrt{(m-m_{\rm M})^2+p^2}$]{\includegraphics[width=0.4\textwidth]{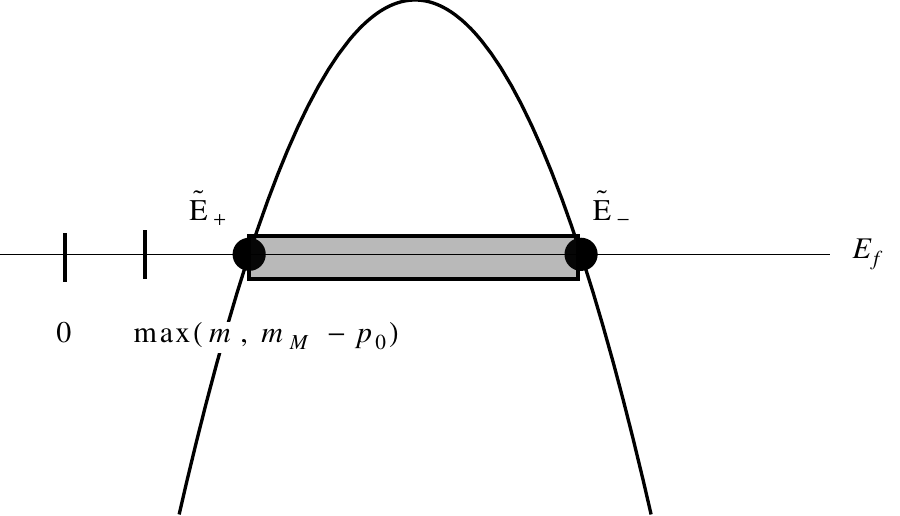}}
  \caption{Summary of integration ranges (gray boxes) for the off-shell imaginary parts of $\Sigma_{0,3,4}(\vec{p},-\i  p_0)$ for Case I. On-shell only the case (b) can be realized.}
  \label{Fig:OffShellSketchesCase1}
\end{center}
\end{figure}

\textbf{Case I.} This is the only case that can be realized on-shell: $E_b=E_f+p_0$. The following two conditions have to be fulfilled: (i) $E_f>m$ and (ii) $E_f>m_{\rm M}-p_0$, which can be summarized in $E_f>\max (m,m_{\rm M}-p_0)$. Evaluating the condition $|\xi|\leq 1$ we find
\begin{equation}
  F(E_f,p,p_0)\geq 0 \;\;\;\;\; \Leftrightarrow \;\;\;\;\; -4s(E_f-\wtEminus)(E_f-\wtEplus)\geq 0\;,
\end{equation}
where we have introduced $s=p_0^2-p^2$ and
\begin{equation}
  \label{DefWidetildeEPlusMinus}
\begin{aligned}
\widetilde{E}_\pm&=-\frac{p_0}{2}+\frac{(m_{\rm M}^2-m^2)p_0}{2s}\\
  &\hspace{0.3cm}\pm\frac{p}{2s}\sqrt{\left[s-(m+m_{\rm M})^2\right]\left[s-(m-m_{\rm M})^2\right]}\;.
\end{aligned}
\end{equation}
These roots of $F(E_f,\cdot,\cdot)$ are generalizations of $E_{\rm max,min}$ introduced in Eq.~\eqref{Eminmax}. One finds indeed
\begin{equation}
\label{OnShellLimitOfTildeEPlusMinus}
  \left.\widetilde{E}_\pm\right|_{s=p_0^2-p^2=m^2>0} = E_{\rm max,min}\;.
\end{equation}
Note that in contrast to $m<E_{\rm min}<E_{\rm max}$, the off-shell roots are not ordered that simply. Dependent on $s>0$ or $s<0$ one has $\wtEminus<\wtEplus$ or $\wtEplus<\wtEminus$, respectively. In addition, it might happen that one or even both roots are negative as we will see.

First, we consider the case $s<0$ which leads to a convex-up parabola $F(E_f,\cdot,\cdot)$ with possible integration ranges $E_f<\wtEplus$ and $E_f>\wtEminus$. In general one has to distinguish additionally the two cases $m<m_{\rm M}$ and $m>m_{\rm M}$, but right now we find for both cases
\begin{equation}
  \wtEplus < -m_{\rm M}-p_0 <0 \;, \;\;\;\;\; \wtEminus > \max (m, m_{\rm M-p_0})\;.
\end{equation}
For Case I with $s<0$ we have the range of integration $E_f>\wtEminus$ as sketched in Fig.~\ref{Fig:OffShellSketchesCase1}(a).

Now consider the case $s>0$ with a concave-down parabola $F(E_f,\cdot,\cdot)$. The possible integration range is $\wtEminus<E_f<\wtEplus$. This time, the roots are not automatically real numbers, but for $(m-m_{\rm M})^2<s<(m+m_{\rm M})^2$ they become purely imaginary and have to be excluded. The first option $p_0<\sqrt{(m-m_{\rm M})^2+p^2}$ leads to
\begin{equation}
\begin{aligned}
  m<m_{\rm M}: & \;\;\;\;\; \wtEplus > \wtEminus > \max(m,m_{\rm M}-p_0)\;,\\
  m>m_{\rm M}: & \;\;\;\;\; \wtEminus < \wtEplus < -  m - p_0 < 0\;.
\end{aligned}
\end{equation}
Therefore, the case $m>m_{\rm M}$ cannot be realized and only for $m<m_{\rm M}$ the full range of integration is accessible. We summarize this case in Fig.~\ref{Fig:OffShellSketchesCase1}(b). Having $s>0$ there is the second option $p_0>\sqrt{(m+m_{\rm M})^2+p^2}$ for which one has for both cases $m<m_{\rm M}$ and $m>m_{\rm M}$:
\begin{equation}
  \wtEplus < -m\;, \;\;\;\;\; \wtEminus>m_{\rm M}-p_0\;.
\end{equation}
We conclude $m_{\rm M}-p_0<\wtEminus<\wtEplus<-m<0$, hence this option is excluded and the discussion of Case I is completed.

\begin{figure}[t!]
\begin{center}
  \subfigure[$s<0$]{\includegraphics[width=0.4\textwidth]{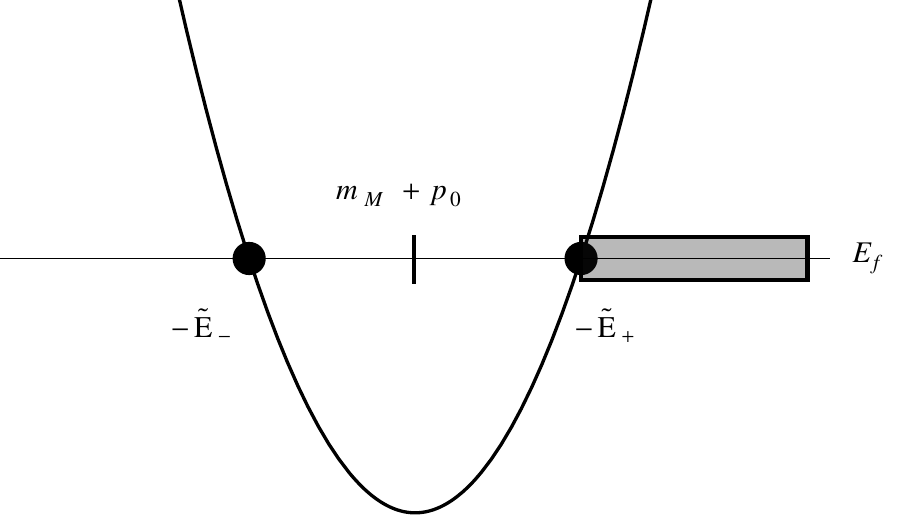}}
  \subfigure[$s>0$ with \mbox{$m>m_{\rm M}$} and $p_0<\sqrt{(m-m_{\rm M})^2+p^2}$]{\includegraphics[width=0.4\textwidth]{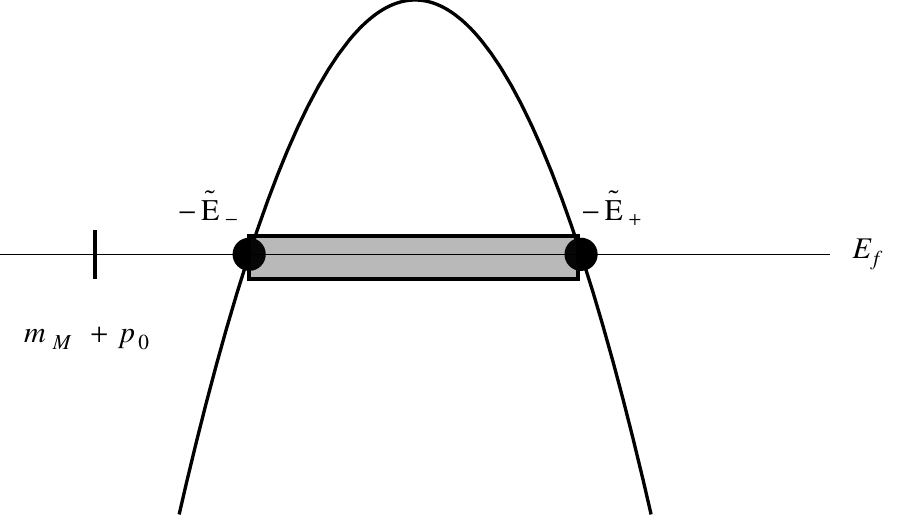}}
  \caption{Summary of integration ranges (gray boxes) for the off-shell imaginary parts of $\Sigma_{0,3,4}(\vec{p},-\i  p_0)$ for Case II.}
  \label{Fig:OffShellSketchesCase2}
\end{center}
\end{figure}

\textbf{Case II.} Evaluating the condition $|\xi|\leq 1$ using $E_b=E_f-p_0$ leads to
\begin{equation}
   F(E_f,p,p_0)\geq 0 \;\;\;\;\; -4s(E_f+\wtEminus)(E_f+\wtEplus)\geq 0\;,
\end{equation}
hence $-\widetilde{E}_\pm$ are the roots of $F(E_f,\cdot,\cdot)$. We follow the same systematic path as before:

Consider first the case $s<0$ implying again a convex-up parabola with $\wtEminus>\wtEplus$. This means $-\wtEminus<-\wtEplus$, providing two possible integration ranges $E_f<-\wtEminus$ and $E_f>-\wtEplus$. One finds:
\begin{equation}
\begin{aligned}
  m<m_{\rm M}: & \;\;\; -\wtEplus > m_{\rm M}+p_0\;, \\
               & \;\;\; -\wtEminus<-m_{\rm M}+p_0<m_{\rm M}+p_0\;,\\
  m>m_{\rm M}: & \;\;\; -\wtEplus > m_{\rm M}+p_0\;,  \\
               & \;\;\; -\wtEminus<-m+p_0<-m_{\rm M}+p_0<m_{\rm M}+p_0\;.
\end{aligned}
\end{equation}
In conclusion we find the range of integration as shown in Fig.~\ref{Fig:OffShellSketchesCase2}(a), again without any restriction on the quark and meson masses.

Now consider $s>0$. This time the possible range of integration is \mbox{$-\wtEplus<E_f<-\wtEminus$}. For the option $p_0<\sqrt{(m-m_{\rm M})^2+p^2}$ we find
\begin{equation}
\begin{aligned}
  m<m_{\rm M}: & \;\;\;\;\; -\wtEplus < -\wtEminus < -m_{\rm M}+p_0 < m_{\rm M}+p_0 \;,\\
  m>m_{\rm M}: & \;\;\;\;\; -\wtEminus > -\wtEplus > m_{\rm M}+p_0\;.
\end{aligned}
\end{equation}
Using the constraint $E_f>m_{\rm M}+p_0$ the case $m<m_{\rm M}$ is excluded and only $m>m_{\rm M}$ is possible. The option $p_0>\sqrt{(m+m_{\rm M})^2+p^2}$, for both cases $m<m_{\rm M}$ and $m>m_{\rm M}$, leads to:
\begin{equation}
  -\wtEplus < -\wtEminus < -m_{\rm M}+p_0<m_{\rm M}+p_0\;,
\end{equation}
which excludes this case because $E_f>\max(m,m_{\rm M}+p_0)$ must be ensured. This case is illustrated in Fig.~\ref{Fig:OffShellSketchesCase2}(b).

\begin{figure}[t!]
\begin{center}
  \subfigure[$s>0$ with \mbox{$p_0>\sqrt{(m+m_{\rm M})^2+p^2}$}]{\includegraphics[width=0.4\textwidth]{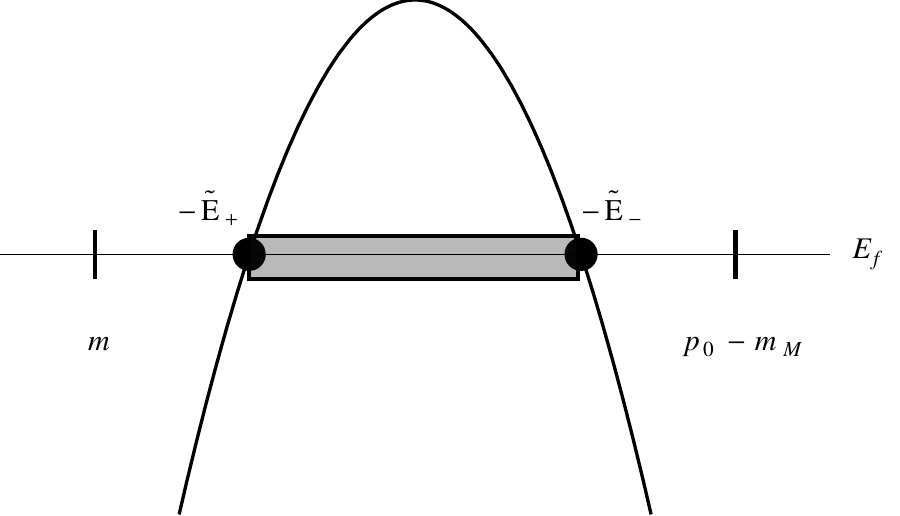}}
  \caption{Summary of integration ranges (gray boxes) for the off-shell imaginary parts of $\Sigma_{0,3,4}(\vec{p},-\i  p_0)$ for Case III.}
  \label{Fig:OffShellSketchesCase3}
\end{center}
\end{figure}

\textbf{Case III.} The final case, $E_b=p_0-E_f$, leads to the two conditions (i) $E_f>m$ and (ii) $E_f<p_0-m_{\rm M}$. From this we get $p_0>m+m_{\rm M}$. It is important to realize that $E_b$ in this case is just the negative of the condition used in Case I. Therefore, all contributions present for Case I cannot be realized for Case III. It remains to check the case $s>0$ in combination with $p_0>\sqrt{(m+m_{\rm M})^2+p^2}$. We have $-\wtEplus < E_f < -\wtEminus$ as possible integration range and find (cf. the related discussion for Case I):
\begin{equation}
  -\wtEplus > m\;, \;\;\;\;\; -\wtEminus <  p_0  -m_{\rm M}\;,
\end{equation}
which is valid for both $m<m_{\rm M}$ and $m>m_{\rm M}$. In conclusion, there is only one contribution to the imaginary part for Case III as shown in Fig.~\ref{Fig:OffShellSketchesCase3}.

Combining now all three cases, the off-shell imaginary part of $\Sigma_0(p,-\i p_0)$ can be calculated immediately. The rather lengthly result reads\footnote{The minus signs for Case I is due to the pole description$+\i\epsilon$ instead of $-\i\epsilon$ for Case II and Case III. For $E_b$ we have always inserted the corresponding relations to $E_f$ and $p_0$ as defined in Eq.~\eqref{CasesDef}.}
\begin{equation}
  \label{DefOffShellJ123}
\begin{aligned}
  \Im\,\Sigma_0^{\rm off} &=\frac{g_{\rm Mqq}^2}{16\pi p} \left\{ \int_{\rm I}\dint E_f\left[-n_{\rm B}(E_f+p_0)-n_{\rm F}^-(E_f)\right]\right.\\
  & \hspace{0.5cm}+\int_{\rm II}\dint E_f\left[n_{\rm B}(E_f-p_0)+n_{\rm F}^+(E_f)\right]\\
  & \hspace{0.5cm}\left. +\int_{\rm III}\dint E_f\left[1+n_{\rm B}(p_0-E_f)-n_{\rm F}^+(E_f)\right]\right\}\\
  & =\frac{g_{\rm Mqq}^2}{16\pi p}\left(J^{\rm I}+J^{\rm II}+J^{\rm III}\right),
\end{aligned}
\end{equation}
with
\begin{equation}
\begin{aligned}
  J^{\rm I} &=\theta(p-p_0)
  \left[\mu-p_0+T\,\ln\frac{n_{\rm F}^-(\wtEminus)}{n_{\rm B}(\wtEminus+p_0)}\right]\\
  &\hspace{0.5cm}+\theta(p_0-p)\theta(m_{\rm M}-m)\theta(\sqrt{(m-m_{\rm M})^2+p^2}-p_0)\,\\
  &\hspace{0.5cm}\times T\,\ln\frac{n_{\rm F}^-(\wtEminus)\,n_{\rm B}(\wtEplus+p_0)}{n_{\rm F}^-(\wtEplus)\,n_{\rm B}(\wtEminus+p_0)}\;,
\end{aligned}
\end{equation}
\begin{equation}
\begin{aligned}
  J^{\rm II} &=\theta(p-p_0)
  \left[\mu-p_0+T\,\ln\frac{n_{\rm B}(-\wtEplus-p_0)}{n_{\rm F}^+(-\wtEplus)}\right]\\
  & +\theta(p_0-p)\theta(m-m_{\rm M})\theta(\sqrt{(m-m_{\rm M})^2+p^2}-p_0)\, \\
  &\times T\,\ln\frac{n_{\rm F}^+(-\wtEminus)\,n_{\rm B}(-\wtEplus-p_0)}{n_{\rm F}^+(-\wtEplus)\,n_{\rm B}(-\wtEminus-p_0)},
\end{aligned}
\end{equation}
\begin{equation}
\begin{aligned}
  J^{\rm III} &=\theta(p_0-\sqrt{(m+m_{\rm M})^2+p^2})\,\\
  &\hspace{0.5cm}\times T\,\ln\frac{n_{\rm F}^-(\wtEplus)\,n_{\rm B}(\wtEminus+p_0)}{n_{\rm F}^-(\wtEminus)\,n_{\rm B}(\wtEplus+p_0)}\;.
\end{aligned}
\end{equation}
Note that for $J^{\rm III}$ the condition $\theta(p_0-p)$ just follows from $\theta(p_0-\sqrt{(m+m_{\rm M})^2+p^2})$, therefore this $\theta$-function can be omitted.

Next we present the off-shell result for the imaginary part of $\Sigma_4(p,-\i p_0)$, performing again a partial-fraction decomposition of Eq.~\eqref{ThermalSigma4AfterMatsubara}:
\begin{equation}
\begin{aligned}
  \label{ThermalSigma4AfterMatsubaraPartialFraction}
\Sigma_4(p,-\i p_0) &= g_{\rm Mqq}^2\int\frac{\dint^3 q}{(2\pi)^3} \frac{1}{4p_0E_bE_f}\\
&\hspace{-1.5cm}\times\left[
\frac{(1-n_{\rm F}^-(E_f)+n_{\rm B}(E_b))(E_b+p_0)}{E_f+E_b+p_0}\right.\\
&\hspace{-1cm}+\frac{(n_{\rm B}(E_b)+n_{\rm F}^-(E_f))(p_0-E_b)}{E_f-E_b+p_0+\i\eps}\\
&\hspace{-1cm}+\frac{(n_{\rm B}(E_b)+n_{\rm F}^+(E_f))(p_0-E_b)}{E_f-E_b-p_0-\i\eps}\\
&\hspace{-1cm}\left.+\frac{(1+n_{\rm B}(E_b)-n_{\rm F}^+(E_f))(p_0-E_b)}{E_f+E_b-p_0-\i\eps}\right].
\end{aligned}
\end{equation}
In comparison to Eq.~\eqref{ThermalSigma03AfterMatsubara} there is the factor $p_0$ in the denominator and also combinations of $E_b$ and $p_0$ in the numerators, but $\Sigma_4$ features the very same pole structure as discussed before. Therefore we find immediately:
\begin{equation}
  \label{DefOffShellK123}
\begin{aligned}
  \Im\,\Sigma_4^{\rm off} &=\frac{g_{\rm Mqq}^2}{16\pi p\,p_0} \\
  & \hspace{-0.2cm}\times\left\{ \int_{\rm I}\dint E_f\,(-E_f)\left[-n_{\rm B}(E_f+p_0)-n_{\rm F}^-(E_f)\right]\right.\\
  & +\int_{\rm II}\dint E_f\, E_f\left[n_{\rm B}(E_f-p_0)+n_{\rm F}^+(E_f)\right]\\
  & \left. +\int_{\rm III}\dint E_f\, E_f\left[1+n_{\rm B}(p_0-E_f)-n_{\rm F}^+(E_f)\right]\right\}\\
  & \hspace{-0.5cm}=\frac{g_{\rm Mqq}^2}{16\pi p\,p_0}\left(K^{\rm I}+K^{\rm II}+K^{\rm III}\right).
\end{aligned}
\end{equation}
Introducing the two auxiliary functions
\begin{equation}
\label{DefAuxiliaryFctG}
  \mathcal{G}^\pm(E)=T^2\left[\frac{\pi^2}{3} + {\rm Li}_2\left(1-\e^{\beta(E\pm p_0)}\right) + {\rm Li}_2\left(-\e^{\beta(E\pm \mu)}\right)\right],
\end{equation}
we find:
\begin{equation}
\begin{aligned}
  K^{\rm I} &= \theta(p-p_0)\left\{\frac{1}{2}(\mu^2-p_0^2)-\wtEminus T\,\ln n_{\rm F}^-(\wtEminus)\right.\\
  &\hspace{2cm}\left.-p_0T\,\ln n_{\rm B}(\wtEminus+p_0) + \mathcal{G}^+(\wtEminus) \right\}\\
  &+\theta(p_0-p)\theta(m_{\rm M}-m)\theta(\sqrt{(m-m_{\rm M})^2+p^2}-p_0)\,\\
  &\times\left\{ p_0 T\;\ln\frac{n_{\rm B}(\wtEplus+p_0)}{n_{\rm B}(\wtEminus+p_0)}+\left(\mathcal{G}^+(\wtEminus)-\mathcal{G}^+(\wtEplus)\right)\right. \\
  & \hspace{0.6cm}\left. + \wtEplus T\,\ln n_{\rm F}^-(\wtEplus) - \wtEminus  T\, \ln n_{\rm F}^-(\wtEminus) \right\}\;,
\end{aligned}
\end{equation}
\begin{equation}
\begin{aligned}
  K^{\rm II} &= \theta(p-p_0)\left\{\frac{1}{2}(\mu^2-p_0^2)+\wtEplus T\,\ln n_{\rm F}^+(-\wtEplus)\right. \\
  &\left. \hspace{1.5cm} +p_0T\,\ln n_{\rm B}(-\wtEplus-p_0) + \mathcal{G}^-(-\wtEplus) \right\} \\
  &\hspace{-0.5cm}+\theta(p_0-p)\theta(m-m_{\rm M})\theta(\sqrt{(m-m_{\rm M})^2+p^2}-p_0)\,\\
  &\hspace{-0.5cm}\times\left\{ p_0 T\;\ln\frac{n_{\rm B}(-\wtEplus-p_0)}{n_{\rm B}(-\wtEminus-p_0)}+\left(\mathcal{G}^-(-\wtEplus)-\mathcal{G}^-(-\wtEminus)\right) \right. \\
  & \hspace{0.1cm}\left. + \wtEplus T\,\ln n_{\rm F}^+(-\wtEplus) - \wtEminus T\, \ln n_{\rm F}^+(-\wtEminus) \right\}\;,
\end{aligned}
\end{equation}
\begin{equation}
\begin{aligned}
  K^{\rm III} &= \theta(p_0-\sqrt{(m+m_{\rm M})^2+p^2})\,\\
  &\hspace{-0.5cm}\times\left\{ -p_0 T\;\ln\frac{n_{\rm B}(\wtEplus+p_0)}{n_{\rm B}(\wtEminus+p_0)}-\left(\mathcal{G}^+(\wtEminus)-\mathcal{G}^+(\wtEplus)\right) \right. \\
  &  \left. + \wtEminus T\,\ln n_{\rm F}^-(\wtEminus) - \wtEplus T\, \ln n_{\rm F}^-(\wtEplus) \right\}\;. \hspace{2.5cm}
\end{aligned}
\end{equation}
Note that the expression for $K^{\rm III}$ is just the negative of the second contribution to $K^{\rm I}$.

Having derived results for $\Im\,\Sigma_0^{\rm off}$ and  $\Im\,\Sigma_4^{\rm off}$, the remaining integral for  $\Sigma_3$ can be performed easily, since all building blocks have been prepared. The main observation is that $\mathcal{F}_3$ splits into two parts: the first one is independent of $E_f$, the second one introduces the same $E_f$ dependence present in the calculation for $\Sigma_4$:
\begin{equation}
\begin{aligned}
 2p^2\mathcal{F}_3&=m_{\rm M}^2+p^2+q^2-E_b^2\\
  &\hspace{-0.8cm}=
  \begin{cases}
  m_{\rm M}^2+p^2-p_0^2-m^2-2E_f p_0 & \mbox{for Case I}\;,\\
  m_{\rm M}^2+p^2-p_0^2-m^2+2E_f p_0 & \mbox{for Case II, III}\;.
  \end{cases}
\end{aligned}
\end{equation}

We find therefore
\begin{equation}
  \label{OffShellIm3imme}
\begin{aligned}
  \Im\,\Sigma_3^{\rm off}=\frac{g_{\rm Mqq}^2}{16\pi p} & \left\{ \int_{\rm I}\dint E_f\,\mathcal{F}_3^{\rm I}\left[-n_{\rm B}(E_f+p_0)-n_{\rm F}^-(E_f)\right]\right.\\
  & \hspace{-1cm}+\int_{\rm II}\dint E_f\,\mathcal{F}_3^{\rm II,III}\left[n_{\rm B}(E_f-p_0)+n_{\rm F}^+(E_f)\right]\\
  & \hspace{-1cm}\left. +\int_{\rm III}\dint E_f\,\mathcal{F}_3^{\rm II,III}\left[1+n_{\rm B}(p_0-E_f)-n_{\rm F}^+(E_f)\right]\right\}\;.
\end{aligned}
\end{equation}
Inspecting the definitions for $J^{\rm x}$ in Eq.~\eqref{DefOffShellJ123} and $K^{\rm x}$ in Eq.~\eqref{DefOffShellK123}, we find the relation:
\begin{equation}
  \Im\,\Sigma_3^{\rm off}=\frac{m_{\rm M}^2+p^2-p_0^2-m^2}{2p^2}\,\Im\,\Sigma_0^{\rm off}+\frac{p_0^2}{p^2}\,\Im\,\Sigma_4^{\rm off}\;.
\end{equation}
~

\end{document}